\documentclass[11pt]{article}
\usepackage[utf8x]{inputenc}
\usepackage[english]{babel}
\usepackage{amsmath,amssymb}
\usepackage{graphicx}
\usepackage{multicol}
\usepackage{xspace}
\usepackage[a4paper,nohead,twoside]{geometry}

\usepackage[bookmarks=true,
           bookmarksopen=true,hidelinks]{hyperref}

\usepackage{cite}
\usepackage[font=small,format=plain,labelfont=bf,up,textfont=it,up]{caption} 
\usepackage[affil-it,max2]{authblk}

\hypersetup{%
  pdftitle = {Numerical solutions of Einstein's equations for cosmological spacetimes with spatial topology S3 and symmetry group U(1)},
  pdfsubject = {},
  pdfauthor = {F. Beyer, L. Escobar and J. Frauendiener},
  pdfkeywords = {}%
}

\numberwithin{equation}{section}

\newcommand{\Eqref}[1]{Eq.~\eqref{#1}}
\newcommand{\Eqsref}[1]{Eqs.~\eqref{#1}}
\newcommand{\Sectionref}[1]{Section~\ref{#1}}
\newcommand{\Sectionsref}[1]{Sections~\ref{#1}}

\newcommand{\Figref}[1]{Fig.~\ref{#1}}

\newcommand{\s}{\hspace{0.1cm}}

\newcommand{\R}{\ensuremath{\mathbb R}\xspace}

\newcommand{\So}{\ensuremath{\mathbb{S}^1}\xspace}
\newcommand{\St}{\ensuremath{\mathbb{S}^2}\xspace}
\newcommand{\Sth}{\ensuremath{\mathbb{S}^3}\xspace}
\newcommand{\U}{\ensuremath{\mathrm{U(1)}}\xspace}

\newcommand{\SU}{\ensuremath{\mathrm{SU(2)}}\xspace}
\newcommand{\mbar}{\overline{m}}

\newcommand{\df} {\mathrm{d}}
\newcommand*{\ii}{\mathrm{i}}

\title{Numerical solutions of Einstein's equations for cosmological spacetimes with spatial topology \Sth and symmetry group \U } 
\author[]{F.\ Beyer\footnote{Email: fbeyer@maths.otago.ac.nz.} }
\author[]{L.\ Escobar\footnote{Email: lescobar@maths.otago.ac.nz.} }
\author[]{J.\ Frauendiener\footnote{Email: joergf@maths.otago.ac.nz.} }
\affil[]{Department of Mathematics and Statistics, University of
  Otago}
\date{\today}

\graphicspath{{.}{plots/}}

\begin{document}

\maketitle

\begin{abstract}
  In this paper we consider the single patch pseudo-spectral scheme for tensorial and spinorial evolution problems on the $2$-sphere presented in \cite{Beyer:2014bu,Beyer:2015bv} which is based
  on the spin-weighted spherical harmonics transform. We apply and extend this method to Einstein's equations and certain classes of spherical cosmological spacetimes. More specifically, we use the 
  hyperbolic reductions of Einstein's equations obtained in the generalized wave map gauge
  formalism combined with Geroch's symmetry reduction, and focus on
  cosmological spacetimes with spatial $\Sth$-topologies and symmetry groups \U or $\U\times\U$. We
  discuss analytical and numerical issues related to our
  implementation. We test our code by reproducing the exact
  inhomogeneous cosmological solutions of the vacuum Einstein field
  equations obtained in \cite{Beyer:2014vw}.
\end{abstract}


\section{Introduction}\label{sec:Introduction}
For many interesting problems, in particular in general relativity,
spherical topologies \St or
\Sth play an important role.  In the context of cosmological models
the spherical
Friedman-Robertson-Walker models, the Bianchi IX or the
Kantowski-Sachs models are particularly important examples. The main difficulty for the numerical (and
analytical) treatment of spherical manifolds is the fact that these manifolds cannot be
covered globally by a single regular coordinate patch, and therefore the
coordinate description of any smooth tensorial quantity inevitably breaks
down somewhere. In the literature this problem is often referred to as
the \textit{pole problem} since in standard polar coordinates for the
$2$-sphere \St these issues appear at the
poles. Many approaches have been tried to deal with this issue, see
for instance \cite{Garfinkle:1999ix,Lehner:2005hc} and references therein. In
earlier work \cite{Beyer:2014bu,Beyer:2015bv}, we have presented a numerical framework
which applies to situations which
involve the $2$-sphere. The main idea of this approach is to implement
and extend the algorithm introduced by Huffenberger and Wandelt (HWT)
in \cite{Huffenberger:2010hh} to compute a transform for functions of
given \textit{spin-weight} $s$ on the $2$-sphere in terms of
\textit{spin-weighted spherical harmonics}. The concepts of the
spin-weight, the so-called \textit{eth}-operators and of
{spin-weighted spherical harmonics} were introduced originally in
\cite{newman1966note} and shall be reviewed in \Sectionref{Sec:swsh}
below.  As a consequence of this formalism, our code is
(pseudo-)spectral in space; time evolutions are performed with the
method of lines and standard ODE integrators (see below). We also point the reader to alternative
implementations of this and similar formalisms in
\cite{gomez1997eth,Beyer:2009vw,Brugmann:2013kt}.

In our earlier work \cite{Beyer:2014bu,Beyer:2015bv}, we have studied simple evolution problems,
like the $2+1$-Maxwell and $2+1$-Dirac equation, on fixed $\St$-backgrounds
as test applications for our numerical infrastructure. The main motivation for this paper now is to
apply the same formalism and numerical infrastructure to the much more
complicated situation of the full Einstein equations. In this context, $2$-spheres
arise in a very natural way.  In the
asymptotically flat setting for example the spatial
manifold $\R\times\St$ is often considered which allows to address the spherical
character of the far zone of the radiation field. 
In the cosmological setting, which we are interested
in here, we can find \St-topologies as a consequence of Geroch's
symmetry reduction \cite{Geroch:1971ix}
(see \Sectionref{sec:gerochReduction}) when the original spatial
manifold has a symmetry.  
For example this was the basis for the
work by Moncrief in \cite{Moncrief:1986th} and for subsequent papers,
and for the work in \cite{Beyer:2011uz,Beyer:2014vw} which shall play
a particularly important role in \Sectionref{sec:Application_test}. Here the spacetimes
of interest have spatial \Sth-topologies and the metrics have a
certain spacelike symmetry such that Geroch's reduction yields the
spatial manifold \St.  The
$3+1$-vacuum Einstein equations thereby become $2+1$-coupled
Einstein-scalar field equations. All of this is explained
in  \Sectionref{sec:gerochReduction}.
Notice that Geroch's symmetry reduction
has also been used to obtain axially symmetric reductions
of Einstein's equations in the asymptotically flat case; see for example
\cite{choptuik2003axisymmetric,Rinne:2010ew}.

The extraction of suitable evolution and constraint equations from Einstein's
equations is essentially the same problem, both in the original $3+1$ and in the
reduced $2+1$-case. We use the \textit{generalized wave map formalism}
\cite{Friedrich:1991nn} (also called {generalized harmonic map formalism} or
simply wave/harmonic map formalism) which can be understood as a covariant
version of the more familiar \textit{generalized wave/harmonic formalism}; the
latter was first introduced in \cite{Friedrich:1991nn} in order to generalize
the original harmonic/wave gauge considered in \cite{FouresBruhat:1952ji}. We
summarize the wave map formalism in \Sectionref{sec:Hyperbolicreduction}. It
turns out that in combination with the spin-weight formalism all singular terms
caused by the singular polar coordinate chart of the $2$-sphere can be completely
eliminated.
This had already been observed for the simpler equations considered in
\cite{Beyer:2014bu,Beyer:2015bv}.
Notice that generalized wave gauges have been used extensively
in the literature in various contexts, see for instance
\cite{Garfinkle:2002ft}.

The numerical results in this paper are obtained
using the spin-weighted spherical harmonics transform in \cite{Beyer:2014bu,Beyer:2015bv}. The underlying Fourier transform is
$2$-dimensional as it applies to functions defined on the
$2$-dimensional manifold $\St$. 
Sometimes however it is interesting to restrict to special classes of functions on $\St$ and therefore to derive a specialized, but more efficient version of this transform. In our context we shall be interested in functions on \St which
are invariant under rotations around an axis (in $\R^3$), i.e., functions which do
not depend on the azimuthal angle $\varphi$ in standard polar
coordinates. For such functions, the $2$-dimensional
transform is inefficient.  In this paper, we therefore also present
an efficient implementation of a $1$-dimensional variant of this
transform which applies to such axially symmetric functions on \St. The complexity
$\mathcal{O}(L^3)$  of the $2$-dimensional transform is thereby reduced to the complexity $\mathcal{O}(L^2)$, where $L$ is the  band limit of the functions on \St in terms of the spin weight spherical harmonics. We call this transform the
\textit{axially symmetric spin-weighted transform}. It will be discussed in detail in \Sectionref{sec:numerical_implementation}.

Finally, \Sectionref{sec:Application_test} is devoted to a test application of
our numerical approach. We discuss different error sources and how they arise in
our implementation. We also study the evolution using the areal gauge and the
generalized wave map gauge.

\section{Preliminaries}\label{sec:Geometric_preliminaries}

\subsection{Geroch's symmetry reduction }\label{sec:gerochReduction}
In this section, we give a quick overview of Geroch's symmetry reduction \cite{Geroch:1971ix}. 
Let $M = \mathbb{R} × \Sigma $ be a globally hyperbolic $4$-dimensional spacetime endowed with a metric $g_{ab}$ of signature ($−, +, +, +$) and  a global smooth time function $t$ whose level sets are Cauchy surfaces homeomorphic to $\Sigma$. We denote the  hypersurfaces given by $t = t_0$ for any constant $t_0$ by $\Sigma_{t_0}$. Each $ \Sigma_{t_0} $ is homeomorphic to $\Sigma$. To begin with, let $\xi^a$ be a smooth space-like Killing vector field on $M$ induced by the smooth effective global action of the group \U. We suppose that $\xi^a$ is everywhere tangent to the hypersurfaces $\Sigma_{t}$ and define
\begin{equation}\label{eq:def_norm_and_twist}
\tilde{\psi} := g_{ab}\xi^a\xi^b, \quad \tilde{\Omega}_a := \epsilon_{abcd} \: \xi^{b} \mathfrak D^c \xi^{d}
\end{equation}
 as the \textit{norm} and the \textit{twist} of $\xi^a$ respectively. The operator $\mathfrak D$ is the covariant derivative compatible with the metric $g_{ab}$. Notice that by the Frobenius theorem, the field $\xi^a$ is hypersurface orthogonal if and only if $\tilde{\Omega}_a = 0$. 
We also define  
\begin{equation}\label{ec:21metric}
\tilde{h}_{ab} := g_{ab} − \frac{1}{\tilde{\psi}} \xi_a \xi_{b}.
\end{equation}
It turns out that for vacuum spacetimes $(M, g_{ab})$ with any cosmological constant $\Lambda$ the $1$-form $\tilde\Omega_a$ is closed. This fact allows us to introduce a local twist potential $\tilde\omega$ so that $\tilde\Omega_a = \mathfrak D_a\tilde\omega$. In fact, $\tilde\omega$ is a \textit{global} potential if $M$ is simply connected as we shall always assume.

Let $S$ be the set of orbits of $\xi^a$ on $M$ and
consider the map
\begin{equation*}
\pi  : M \to S ,
\end{equation*}
where  $\pi$ maps every $p \in M$ to the uniquely determined integral curve of $\xi^a$ through $p$. The requirement that $\pi$ is a smooth map induces a differentiable structure on $S$, and hence it can be considered as a smooth manifold. Since $\mathcal L_{\xi} \tilde h_{ab}=0$, there is a unique smooth Lorentzian metric on $S$ which pulls back to $\tilde{h}_{ab}$ along $\pi$. We write this metric on $S$  as $\hat{h}_{ab}$. For the same reason, there are also unique functions $\psi$, $\omega$ on $S$ which  pull back to $\tilde{\psi}$ and $\tilde\omega$.
It turns out that Einstein's field equations with cosmological constant $\Lambda$ for $(M,g_{ab})$ imply
the following set of equations for $(S, h_{ab}, \psi,\omega)$ where
\begin{equation}\label{eq:conformal_metric}
h_{ab} := \psi \hat{h}_{ab}.
\end{equation}
We call this system  \textit{the Geroch-Einstein system} (GES)\footnote{We have used $\nabla_{a}$ to denote the covariant derivative operator associated with $h_{ab}$. Indices are lowered and raised by $h_{ab}$.} and it reads
\begin{eqnarray}\label{eq:geroch_einstein_equations}
 \nabla_{a} \nabla^{a} \psi   &=& \dfrac{1}{\psi} \left(  \nabla_{a} \psi \nabla^{a} \psi  -  \nabla_{a} \omega \nabla^{a} \omega \right) - 2 \Lambda, \nonumber\\
 \nabla_{a} \nabla^{a} \omega &=& \dfrac{1}{\psi} \nabla^{a} \psi \nabla_{a} \omega , \\
R _{ab} &=&   E_{ab} + \dfrac{2\Lambda}{\psi} h_{ ab }   , \nonumber 
\end{eqnarray}
with
\begin{equation}
  \label{eq:Sab}
  E_{ab}  =  \dfrac{1}{2\psi^2 } \left(  \nabla_{a} \psi \nabla_{b} \psi + \nabla_{a} \omega \nabla_{b} \omega \right) ,
\end{equation}
where $R_{ab}$ is the Ricci tensor associated with $h_{ab}$.
In fact, the equations for $\psi$ and $\omega$ imply that
\begin{equation}\label{eq:energy_momentum_tensor}
T_{ab}:= E_{ab} - \dfrac{1}{2} h_{ab} E, 
\end{equation}
is divergence free. It thus plays the role of the energy momentum tensor associated with the two scalar fields $\psi$ and $\omega$ in the $2+1$-dimensional spacetime $S$ with metric $h_{ab}$. As a result, the GES can be interpreted as the equations of 2+1-gravity coupled to two scalar fields $\psi$ and $\omega$ governed by wave-map equations. 

Suppose for a moment that $\psi$, $\omega$ and ${h}_{ab}$ are known and that they satisfy \Eqref{eq:geroch_einstein_equations}. Then  we can reconstruct a solution $(M, g_{ab})$ of the $3+1$-dimensional Einstein vacuum equation with cosmological constant $\Lambda$ as follows. It turns out that as a consequence of the above equations, the $2$-form
\[\alpha_{ab}=\frac{1}{2\psi^{3/2}} \epsilon_{abc}\Omega^c\]
on $S$
is curl-free, i.e.,
\[\nabla_{[a}\alpha_{bc]}=0.\]
We can pull this quantity back to a $2$-form $\tilde\alpha$ on $M$ which is also curl-free. This means that there exists, locally, a $1$-form $\tilde\eta_a$ on $M$ such that
\[\mathfrak D _{[a}\tilde\eta_{b]}=\tilde\alpha_{ab};\]
notice that it is irrelevant here that the Levi-Civita covariant derivative $\mathfrak D _a$ is not yet known at this stage due to the antisymmetrization. In fact, the $1$-form $\eta_a$ is uniquely determined by this equation up to the gradient of a smooth function $f$ and we can use some of the freedom in choosing $f$ to set $\eta_a\xi^a=1$. Then the metric
\[g_{ab}=\tilde h_{ab}+\tilde\psi \tilde\eta_a\tilde\eta_b,\]
where $\tilde h_{ab}$ and $\tilde\psi$ are the pull-backs of $h_{ab}/\psi$ and $\psi$, respectively,
is a solution of the $3+1$ Einstein vacuum equation with cosmological constant, $\Lambda$, and, we find
\[\xi_a=g_{ab}\xi^b=\psi\eta_a.\]
So effectively, the quantities $\psi$, $\omega$ and ${h}_{ab}$ determine the metric $g_{ab}$ up to a total gradient of some function $f$ which fixes the covariant version $\xi_a$ of the Killing vector field $\xi^a$.

\subsection{\texorpdfstring{The $2$- and $3$-spheres}{The 2- and 3-spheres}}
\label{subsec:s3topology}
To begin with, we consider the manifold \Sth as the submanifold of $\R^4$ given by $x_1^2+x_2^2+x_3^2+x_4^2=1$. We can introduce \textit{Euler coordinates} $(\theta,\lambda_1,\lambda_2)$ on $\Sth$ 
  \begin{equation*}    
   \begin{split}
     x_1&=\cos\frac\theta 2\cos\lambda_1, 
     \quad x_2=\cos\frac\theta 2\sin\lambda_1,\\
      x_3&=\sin\frac\theta 2\cos\lambda_2, 
      \quad x_4=\sin\frac\theta 2\sin\lambda_2,
    \end{split}
  \end{equation*}
  where $\theta\in (0,\pi)$ and $\lambda_1,\lambda_2\in (0,2\pi)$. 
Alternatively, we use coordinates $(\theta,{\rho_1},{\rho_2})$ (which are also referred to as Euler coordinates) with $\theta$ as above and 
  \begin{equation}\label{eq:coordinates_rho}
    \lambda_1=( \rho_1+\rho_2 )/2,\quad
    \lambda_2=(\rho_1-\rho_2 )/2.
  \end{equation}
Clearly, both sets of Euler coordinates break down at $\theta=0$ and $\pi$. The vector fields $\partial_{\rho_{1}}$ and $\partial_{\rho_{2}}$ are smooth and non-vanishing vector fields on \Sth which become parallel at the poles $\theta = 0,\pi$. 

Similarly we define the manifold \St as the subset $y_1^2+y_2^2+y_3^2=1$ of $\R^3$ and introduce standard polar coordinates 
\[y_1=\sin\vartheta\cos{\varphi},\quad y_2=\sin\vartheta\sin{\varphi},\quad y_3=\cos\vartheta.\]
The \textit{Hopf map} $\pi:\Sth\rightarrow\St$ is
\begin{align*}
  (x_1,x_2,x_3,x_4)\mapsto
  (y_1,y_2,y_3)
  &=(2(x_1 x_3+x_2x_4),2(x_2x_3-x_1x_4),
  x_1^2+x_2^2-x_3^2-x_4^2)\\
  &=(\sin\theta\cos{\rho_2},\sin\theta\sin{\rho_2},\cos\theta).
\end{align*}
This is a smooth map which has the coordinate representation
\begin{equation}\label{eq:localHopf}
  \pi:(\theta,{\rho_1},{\rho_2})\mapsto (\vartheta,\varphi)=(\theta,{\rho_2}).
\end{equation}
Hence, with respect to our coordinates, the Hopf map reduces to a simple
projection map.  Now, $\Sth$ is a principal fiber bundle over $\St$ with
structure group $\U$ whose bundle map is the Hopf map. In fact, if
$M=\R\times\Sth$ and $\xi^a=\partial_{\rho_1}^a$ is assumed to be a Killing
vector (as we will do) then $S=\R\times\St$ is the space of orbits and $\pi$ the
corresponding map in Geroch's symmetry reduction.

In this paper, we shall employ this relationship between \Sth and \St for our studies of $U(1)$-symmetric fields. Just as a side remark we also mention the case $\Sigma=\So\times\St$ which is a trivial bundle over $\St$. If $\xi^a$ agrees with a vector field tangent to the $\So$-factor and we introduce appropriate coordinates, then the bundle map $\pi: \So\times\St\rightarrow\St$ takes the same coordinate form as \Eqref{eq:localHopf}. In particular, Geroch's symmetry reduction also yields  the space of orbits $S=\R\times\St$. Hence, almost all techniques which we introduce in this paper, can also be applied to that case.

\subsection{The bundle of orthonormal frames over \texorpdfstring{$\St$}{S2} and
  spin-weighted spherical harmonics}
\label{Sec:swsh}

$\mathrm{SO}(3)$ is the bundle of oriented orthonormal frames over \St with
structure group $U(1)$. Given that $\mathrm{SO}(3)$ is double covered by
$\mathrm{SU}(2)$ and that the latter is diffeomorphic to $\Sth$, the Hopf map
$\pi: \Sth\rightarrow\St$ can be identified with the bundle map. The theoretical
details are discussed, for example, in \cite{Beyer:2014bu}. Hence, when we start
from the spatial manifold $\Sth$, do the symmetry reduction as explained before
and therefore arrive at the spatial manifold $\St$, the manifold $\Sth$
``reappears'' in a different role, namely as the bundle of orthonormal
frames. In practice this means the following: we let $U$ be the dense open subset of $\St$ obtained by removing the north and the south poles. The polar coordinates $(\vartheta,\varphi)$ cover $U$ and the Euler coordinates $(\theta,\rho_1,\rho_2)$ cover $\pi^{-1}(U)$. In particular, \Eqref{eq:localHopf} holds.
Let $(m^a, \overline{m}^a )$ be the complex smooth frame on $U$ defined by
\begin{equation}\label{eq:referenceframe}
m^a:=\frac 1{\sqrt 2}\left(\partial_{\vartheta}^a-\frac \ii{\sin\theta}\partial_\varphi^a\right)
\end{equation}
and by the complex conjugate $\overline{m}^a$. Any point $p=(\theta,\rho_1,\rho_2)\in\Sth$ in the bundle of orthonormal frames can then be identified with the basis $(e^{\ii \rho_1}m^a,e^{-\ii \rho_1}\overline{m}^a)$ of the tangent space evaluated at the point $\pi(p)\in\St$.
The local section $\sigma: U\rightarrow \pi^{-1}(U)$ specified by any real function  $\rho_1=\rho_1(\vartheta,\varphi)$ 
yields a different frame over $U$ which is related to $(m^a,\overline{m}^a)$ by a pointwise rotation
\begin{equation}
  \label{eq:framerotation}
  m^a\mapsto e^{\ii \rho_1(\vartheta,\varphi)}m^a,\quad \overline{m}^a\mapsto e^{-\ii \rho_1(\vartheta,\varphi)}\overline{m}^a
\end{equation}
at each point in $U$.
If $f: U\rightarrow\mathbb C$ is a component of a smooth real tensor field on $\St$ with respect to the frame $(m^a,\overline{m}^a)$, the function $e^{is\rho_1}\cdot (f\circ\pi)$ on $\pi^{-1}(U)\subset \Sth$, which is defined for some integer $s$ called the \textit{spin-weight}, is the corresponding component obtained by any frame rotation above. 
Any such function $f$ is said to have the \textit{well-defined spin-weight}~$s$.
The ``standard'' section and hence the ``standard'' frame which we shall use without further notice in the following is given by $\rho_1(\vartheta,\varphi)=0$. We shall not distinguish between the original function $f$ on $U$ and the corresponding function $f\circ\pi$ on the range of the standard section in the bundle of orthonormal frames.
When we interpret a function $f$ on $\St$ with well-defined spin-weight $s$ as the function $e^{is\rho_1}\cdot (f\circ\pi)$ on $\pi^{-1}(U)\subset \Sth$, we are able to replace \textit{singular} frame derivatives on $\St$ by \textit{regular} derivatives along left-invariant vector fields on \Sth.  This yields \textit{eth}-operators $\eth$ and $\bar{\eth}$ given by
\begin{eqnarray}\label{eq:def_eths}
\eth [f]       &:=& \partial_\vartheta [f] - \dfrac{\ii}{ \text{sin} \vartheta} \partial_\varphi [f] - s f \text{cot} \vartheta  , \\
\bar{\eth} [f] &:=& \partial_\vartheta [f] + \dfrac{\ii}{ \text{sin} \vartheta} \partial_\varphi [f] + s f \text{cot} \vartheta  ,
\end{eqnarray}
for any function $f$ on $\mathbb{S}^2$ with spin-weight $s$. Notice that our convention differs by a factor $\sqrt{2}$ from the one for $m^a$ and $\overline{m}^a$ in \Eqref{eq:referenceframe}. The function $\eth [f]$ has a well-defined spin-weight $s+1$ and $\bar{\eth} [f]$ has spin-weight $s-1$.

The \textit{spin-weighted  spherical harmonics} (SWSH) play an important role in the representation of spin-weighted functions on $\St$.  
They form a basis of $L^2(\SU)$ as an application of  the Peter-Weyl theorem to the compact group $\SU$ \cite{Sugiura:1990vj}. Under certain assumptions, any spin-weighted function ${}_sf$ on $\St$ can therefore be represented as an infinite series of SWSH
\begin{equation*}
\label{eq:function}
{}_sf( \vartheta , \varphi) = \sum\limits_{l=0}^{\infty}  \sum\limits_{m=-l}^{l}
\, a_{lm} \,{}_{s}Y_{lm} (\vartheta,\varphi) ,
\end{equation*}
where $\hspace{0.2cm}_{s}Y_{lm}$ are the SWSH and $a_{lm}$ the complex coefficients of the function (also called spectral coefficients). 
The standard scalar spherical harmonics are given by $s = 0$. 
By applying the $eth$-operators to these we obtain
\begin{eqnarray*}\label{eq:eths}
\eth [ \hspace{0.1cm}_{s}Y_{lm} (\vartheta,\varphi) ] &=& - \sqrt{ (l-s)(l+s+1) } \hspace{0.1cm}_{s+1}Y_{lm} (\vartheta,\varphi) , \nonumber \\
\bar{\eth}  [ \hspace{0.1cm}_{s}Y_{lm} (\vartheta,\varphi) ]  &=& \sqrt{ (l+s)(l-s+1) } \hspace{0.1cm}_{s-1}Y_{lm} (\vartheta,\varphi) , \nonumber\\
\bar{\eth} \eth  [ \hspace{0.1cm}_{s}Y_{lm} (\vartheta,\varphi) ]  &=& - (l-s)(l+s+1) \hspace{0.1cm}_{s}Y_{lm} (\vartheta,\varphi) .
\end{eqnarray*}

\section{Einstein's evolution and constraint equations}\label{sec:Geroch_Einstein_evolution_equation}

\subsection{Hyperbolic reduction}
\label{sec:Hyperbolicreduction}

The Einstein equations~\eqref{eq:geroch_einstein_equations} are a set of
geometric partial differential equations. They are invariant under general
coordinate transformations which implies that they are not automatically of
any particular type when expressed in an arbitrary coordinate system. There
exist many ways of extracting hyperbolic and elliptic subsets from these
equations by fixing certain coordinate gauges. Here, we will use the so called
wave map gauge, a generalization of the well-known harmonic gauge. The setup for
the wave map gauge is discussed in detail in the appendix. 

We now introduce a general smooth frame $(e_\mu^a)$. Notice that this frame is neither necessarily a coordinate frame nor an orthonormal frame. The components $R_{\mu\nu}$ of the Ricci tensor $R_{ab}$ with respect to this frame can be written as
\begin{equation}\label{eq:Ricci1}
R_{\sigma\rho} = -\dfrac{1}{2} \; h^{\mu\nu} \partial_\mu\partial_\nu h_{\sigma\rho} + \nabla_{(\sigma}\Gamma_{\rho)} + \Upsilon_{\sigma\rho}( h, \partial h ),
\end{equation}
where in the first term, $\partial_\nu h_{\sigma\rho}$ is the derivative of the function $h_{\sigma\rho}$ in the direction of the frame vector field $e_\nu^a$,
the third term is a lengthy nonlinear expression in the components of $h_{ab}$ and their first derivatives, and $\Gamma_{\mu}$ in the second term denotes the \textit{contracted connection coefficients} $\Gamma_{\mu}:=h^{\nu\sigma}\Gamma_{\mu\nu\sigma}$ with
\begin{equation*}
\nabla_{\mu} \Gamma_{\nu} := \partial_\mu\Gamma_{\nu} - \Gamma^\sigma{}_{\mu\nu}  \Gamma_{\sigma}.
 \end{equation*}
Here, the connection coefficients of the frame are defined as (using
the conventions in \cite{Stephani:2003uy})
\[
\nabla_\mu e_\nu^a=\Gamma^\sigma{}_{\mu\nu} e_\sigma^a
\]
and are computed as
\begin{equation}\label{ec:connection_coefficients}
\Gamma^\sigma{}_{\mu\nu}=h^{\rho\sigma}\Gamma_{ \rho\mu \nu },\quad \Gamma _{ \mu \nu \rho }=\frac{1}{2}\left(\text{ }\partial_{\rho} h_{ \mu \nu } + \partial_{\nu} h_{ \mu \rho } - \partial_{\mu} h_{ \nu \rho } + C_{ \rho \mu \nu} + C_{ \nu \mu \rho} -C_{ \mu \nu \rho} \text{  } \right),
\end{equation}
with  $C_{ \nu \mu \rho} = h_{\nu\sigma} C^{\sigma}{}_{\mu \rho}$ and
\[ 
[ e_{\mu}, e_{\nu} ]^a =C^{\rho}{}_{\mu\nu}e_{\rho}^a.
\]
Notice that while the left side of \Eqref{eq:Ricci1} represents the components of a smooth tensor field, none of the terms on the right-hand side does this individually. In particular, the quantity $\Gamma_{\mu}$ does not represent a covector field. Notice also that, in general, $\partial_\mu \partial_\nu h_{\sigma\rho}$ is not the same as $\partial_\nu \partial_\mu h_{\sigma\rho}$ (as it would be for a coordinate frame for which $C_{ \nu \mu \rho}=0$).

The non-tensorial split is not the only issue of \Eqref{eq:Ricci1}. In addition, the second term destroys the hyperbolicity of its principal part. The idea is to get rid of this second term by defining a new tensor field 
\begin{equation}
\label{eq:riccihat}
\hat R_{ab}:= R_{ab}+\nabla_{(a} \mathcal{D}_{b)}
\end{equation}
where $\mathcal{D}^a$ is the vector field defined in \Eqref{eq:defDs}. The frame
representation of $\hat R_{ab}$ is
\[
\hat R_{\sigma\rho} = -\dfrac{1}{2} \; h^{\mu\nu} \partial_\mu\partial_\nu
h_{\sigma\rho}   + \Upsilon_{\sigma\rho}(
h, \partial h )  + h_{\alpha(\rho}\,\nabla_{\sigma)} \bar\Gamma^\alpha{}_{\beta\gamma}
h^{\beta\gamma}+ \nabla_{(\sigma} f_{\rho)}, 
\]
where $\Upsilon_{\sigma\rho}$ is the same nonlinear expression as above. Regarded as a differential operator acting on
$h_{\mu\nu}$ it has a hyperbolic principal part.

The idea of the generalized wave map formalism is to replace the Ricci tensor $R_{ab}$ in the field equation by this new tensor $\hat R_{ab}$. We call the resulting equations the ``evolution equations'' since, under suitable conditions, these have a well-posed initial value problem for any choice of \textit{gauge source quantities} $f_a$ and $\bar h_{ab}$. The evolution equations implied by \Eqref{eq:geroch_einstein_equations} are therefore
\begin{eqnarray*}
 \nabla_{a} \nabla^{a} \psi   &=& \dfrac{1}{\psi} \left(  \nabla_{a} \psi \nabla^{a} \psi  -  \nabla_{a} \omega \nabla^{a} \omega \right) - 2 \Lambda, \nonumber\\
 \nabla_{a} \nabla^{a} \omega &=& \dfrac{1}{\psi} \nabla^{a} \psi \nabla_{a} \omega , \\
\hat R _{ab} &=&   E_{ab} + \dfrac{2\Lambda}{\psi} h_{ ab }   , \nonumber 
\end{eqnarray*}
with \Eqsref{eq:Sab} and \eqref{eq:riccihat}. This is a coupled system of
quasilinear wave equations. 

Suppose now that $(h_{ab},\psi,\omega)$ is a solution of the evolution equations. It is a solution of the original equations \eqref{eq:geroch_einstein_equations} if $\mathcal{D}^a\equiv 0$ (hence, if $h_{ab}$ is in generalized wave map gauge) and hence $\hat R_{ab}\equiv R_{ab}$. Under which conditions does therefore the covector field $\mathcal{D}^a$ vanish?
The evolution equations and the contracted Bianchi identities imply
the \textit{subsidiary system} (see \cite{Ringstrom:2009cj} for details) 
\begin{equation}\label{subsidiarysystem}
\nabla_{b} \nabla^{b} \mathcal{D}_{a} + \mathcal{D}^{b}  \nabla_{(b} \mathcal{D}_{a)} = 0.
\end{equation}
This is a homogeneous system of wave equations for the unknown $\mathcal{D}_a$. It follows that $\mathcal{D}^a\equiv 0$ if and only if $\mathcal{D}^a=0$ and $\nabla_a \mathcal{D}^b=0$ on the initial hypersurface; these conditions therefore constitute \textit{constraints}.
While the constraint 
\begin{equation}
\label{eq:constraint1}
  0=\mathcal{D}^\nu=h^{\rho\sigma}(\bar\Gamma^{\nu}{}_{\rho\sigma}-\Gamma^{\nu}{}_{\rho\sigma})+f^\nu
\end{equation}
can be satisfied for any initial data $h_{ab}$, $\psi$ and $\omega$ by a suitable choice of the free gauge source quantities $f_a$ and $\bar h_{ab}$, and is hence referred to as the \textit{gauge constraint}, the constraints
\begin{equation}
  \label{eq:constraint2}
  \nabla_\mu \mathcal{D}_\nu=0
\end{equation}
turn out to hold at the initial time if and only if the initial data satisfy the standard Hamiltonian and Momentum constraints (supposing that the gauge constraint and the evolution equations are satisfied). These are equations which are therefore independent of the gauge source functions. Hence \Eqref{eq:constraint2} represents the actual ``physical constraints'' on the initial data for $h_{ab}$, $\psi$ and $\omega$.

\subsection{The generalized wave map gauge in the case \texorpdfstring{$S=\R\times\St$}{S=RxS2}}\label{Sec:evol_equations}


In this section, we focus on the case $S:=\R\times\St$ and the field equations in the form \Eqref{eq:geroch_einstein_equations}.
As before let $t: S\rightarrow\R$ be a smooth time function on $S$ and
\begin{equation*}
\Sigma_t:=\{t\}\times\St\simeq \St, \quad t\in\R.
\end{equation*}
We introduce coordinates $(t,\vartheta,\varphi)$ on the dense subset $\R\times U$ of $S$ and define $T^a=\partial_t^a$.
With the same choice of complex vector field $m^a$ on $U\subset\St$ as in \Sectionref{subsec:s3topology}, we introduce the frame $(e_0^a,e_1^a,e_2^a)=(T^a,m^a,\mbar^a)$ on $\R\times U$. The spin-weight of any function $f: \R\times U\rightarrow\mathbb C$ is defined in the same way as in \Sectionref{Sec:swsh}, but now with respect to frame transformations of the form
\begin{equation*}
\label{eq:framerotation2}
T^a\mapsto T^a, \quad m^a\mapsto e^{\ii\rho_1(\vartheta,\varphi)} m^a,
\quad \overline{m}^a\mapsto e^{-\ii \rho_1(\vartheta,\varphi)}\mbar^a
\end{equation*}
instead of \Eqref{eq:framerotation}.
Therefore, the frame vector field $T^a$ has spin-weight $0$, $m^a$ spin-weight $1$ and $\mbar^a$ spin-weight $-1$. 
Under the above considerations, we choose the dual frame  $( \omega^0_a ,\omega^1_a, \omega^2_a )$ by
\begin{equation*}
\omega_a^0 = \nabla_a t , \quad  \omega^1_a = \frac 1{\sqrt 2} \left( \nabla_a \vartheta+\ii \sin\vartheta \hspace{0.1cm} \nabla_a \varphi \right), \quad \omega^2_a = \overline{\omega}^1_a,
\end{equation*}
with spin-weight of $0$, $-1$ and $1$ respectively. The duality relation reads
\[\omega_a^\mu e_\nu^a=\delta^\mu_\nu.\]
Then, the general form of a smooth metric on $S$ is 
\begin{equation}\label{3Dmetric_smoothframe}
h_{ab} =   \lambda \hspace{0.1cm} \omega_a^0 \omega_b^0  + 2 \hspace{0.1cm} \omega_{(a}^0 \left(  \beta \hspace{0.1cm} \omega_{b)}^1 + \bar{\beta} \hspace{0.1cm} \omega_{b)}^2  \right)  +  2 \delta \hspace{0.1cm} \omega_{(a}^1 \omega_{b)}^2  + \phi \hspace{0.1cm} \omega_a^1 \omega_b^1 + \bar{ \phi } \hspace{0.1cm} \omega_a^2 \omega_b^2.
\end{equation}
After a straightforward computation we find that almost all the quantities $C^{\mu}{}_{\nu\rho}$ introduced in the previous subsection are zero except
\begin{equation}
\label{eq:structuresphere}
C^2{}_{12}= C^1{}_{21} =- C^2{}_{21} = - C^1{}_{12}= \frac{-1}{\sqrt 2}\cot\vartheta .  
\end{equation}
The occurrence of the singular function $\cot\vartheta$ is a consequence of the
fact that the quantities $C^{\mu}{}_{\nu\rho}$ are not components of a tensor
field and hence do not have well-defined spin-weights. It is a consequence of
the discussion in the previous section, however, that all quantities, which we
\textit{eventually} work with, \textit{are} frame components of smooth tensor
fields and therefore \textit{have} well-defined spin-weights without such
singular terms --- even though singular terms without well-defined spin-weights
appear in intermediate calculations when non-tensorial
expressions are used. Since the eth-operators are essentially projections of 
covariant derivatives it is not surprising that
all these terms which are caused by the connection coefficients related to the
unit sphere will disappear when frame derivatives are replaced
consistently by corresponding eth-operators according to \Eqref{eq:def_eths}. Indeed, we are able to demonstrate this explicitly.

In all of what follows we choose 
\begin{equation}\label{eq:background_metric}
  \bar{h}_{ab} = -\omega_a^0 \omega_b^0  +  2 \hspace{0.1cm} \omega_{(a}^1\omega_{b)}^2 ,
\end{equation}
as the reference metric introduced in the previous section. This is a smooth metric on $S$ which represents the static cylinder with the standard spatial metric on \St. 
All the remaining gauge freedom is then encoded in the vector field $f^a$. We shall introduce
quantities 
\begin{align}
  \label{eq:gammas2}
  \breve{\Gamma}^\mu    &:= 
                  h^{\sigma\rho}\bar{\Gamma}^{\mu}
                             {}_{\sigma\rho},\\
  \label{eq:gammas}
  \mathring{\Gamma}^\mu &:= \Gamma^\mu -  \breve{\Gamma}^\mu,  
\end{align}
where the latter are the components of a covector field $\mathring{\Gamma}_a$ which we call the \textit{smooth contracted connection coefficients}. Thus
\[ \mathcal{D}^{a} =  \mathring{\Gamma}^{a} -  f^{a}.\]
Note that by construction, the non-tensorial quantities $\breve{\Gamma}^\mu$  do not contain any derivatives of the metric $h_{ab}$ and we have
\begin{equation*}
\bar{\Gamma}^{a} {}_{bc}=\frac{ \bar{h}^{ad} }{2}\left(\text{ }  C_{d b c} + C_{d  c b} - C_{b c d} \text{  } \right)
\end{equation*}
as a consequence of \Eqref{ec:connection_coefficients}. But they contain terms proportional to $\cot\vartheta$ due to \Eqref{eq:structuresphere}.
All the first order derivatives of the metric $h_{ab}$ in $\Gamma_\mu$ are in $\mathring{\Gamma}_{a}$. 
We define the non-tensorial quantity
\begin{equation}
\mathring{\Upsilon}_{\mu\nu}( h, \partial h ,  \breve{\Gamma} ) :=  \nabla_{(\mu}  \breve{\Gamma}_{\nu)}  + \Upsilon_{\mu\nu}( h, \partial h ) \;,
\end{equation} 
with the same $\Upsilon_{\mu\nu}$ as in \Eqref{eq:Ricci1}, and write the evolution equations as
\begin{eqnarray}\label{eq:geroch_einstein_equations_final}
h^{\mu\nu} \partial_\mu \partial_\nu \psi -h^{\mu\nu} \Gamma^{\rho} {}_{\nu\mu} \partial_\rho \psi  &=&  \dfrac{h^{\rho\sigma}}{\psi} \left(  \partial_{\rho} \psi \partial_{\sigma} \psi - \partial_{\rho} \omega \partial_{\sigma} \omega \right) - 2 \Lambda , \nonumber\\
h^{\mu\nu} \partial_\mu \partial_\nu \omega -h^{\mu\nu} \Gamma^{\rho} {}_{\nu\mu} \partial_\rho \omega &=&  \dfrac{ h^{\rho\sigma} }{\psi} \: \partial_{\rho} \psi \partial_{\sigma} \omega ,  \\
 h^{\rho\sigma} \partial_\rho \partial_{\sigma} h_{\mu\nu}  - 2 \mathring{\Upsilon}_{\mu\nu}( h, \partial h ,  \breve{\Gamma} )&=&  2 \nabla_{(\mu} f_{\nu)} - \dfrac{1}{\psi^2 } \left(  \partial_{\mu} \psi \; \partial_{\nu} \psi + \partial_{\mu} \omega \; \partial_{\nu} \omega \right) -  \dfrac{4\Lambda}{\psi} h_{ \mu\nu } \:. \nonumber
\end{eqnarray}
We notice that the first terms on the left hand sides constitute the principal part of this evolution system, i.e., quasilinear wave operators. These terms by themselves are not tensorial and hence give rise to singular terms (terms proportional to $\cot\vartheta$) and terms which do not have well-defined spin-weights when the frame derivatives are replaced by eth-operators as described before. The second terms on the left hand sides cancel these problematic terms completely, and consequently, the left hand sides are tensorial. The right hand sides are tensorial already. 
As a result of this fully tensorial character of all these equations, the system of evolution equations \Eqref{eq:geroch_einstein_equations_final} can now be solved by implementing a pseudo-spectral method based on the SWSH.

\section{Numerical implementation}\label{sec:numerical_implementation}

As explained earlier, we wish to implement a spectral method based on spin-weighted spherical harmonics to approximate spatial derivatives. A basic introduction to spectral methods can be found in books like \cite{Durran:2010,Fornberg:1998gv,Vretblad:2003} and references therein. For the temporal discretization we mainly used the Runge-Kutta-Fehlberg method except for convergence tests for which the explicit 4th-order Runge-Kutta method is used.
We start this section by describing briefly the algorithm of complexity $\mathcal{O}(L^3)$, where $L$ is the  band limit of the functions on \St in terms of the spin-weighted spherical harmonics, to compute the spin-weighted spherical harmonic transforms (forward and backward) introduced by Huffenberger and Wandelt in \cite{Huffenberger:2010hh}. Henceforth we will refer to this algorithm as HWT. Later, in the next subsection, we introduce an optimized version of this transform for the case of functions on \St with spin-weight $s$ that exhibit axial symmetry (i.e., invariant along the coordinate vector field $\partial_\varphi$). As a result, we obtain an algorithm of complexity  $\mathcal{O}(L^2)$ which requires a low memory cost in comparison with that required by HWTs. In this work, we will focus on functions  that exhibit axial symmetry and hence  our spectral implementation is based on this transform. For details, improvements and applications of the HWTs for general functions in \St  we refer the reader to \cite{Beyer:2015bv,Beyer:2014bu}. We finalize this section by discussing our method to choose the ``optimal'' grid size in order to keep numerical errors as small as possible.

\subsection{General description of HWTs}\label{sec:spin_transform}
To begin with, let us consider a square integrable spin-weighted function $f \in L^2(\mathbb{S}^2)$ with spin-weight $s$. The forward and backward  spin-weighted spherical harmonic transformations are defined, respectively,  by 
\begin{equation}\label{alm}
{}_s a_{lm} = \int \limits_{\mathbb{S}^2} \s f(\vartheta,\varphi) \s { _s \overline Y_{lm}}(\vartheta,\varphi) \s d\Omega , 
\end{equation}
\begin{equation}\label{eq:fun}
 f(\vartheta,\varphi)= \sum^L_{l=|s|}\sum^l_{m=-l} {{}_sa_{lm}} \,{{}_sY_{lm}(\vartheta,\varphi)},
\end{equation}  
where the decomposition has been truncated at the maximal mode $L$. Henceforth, we shall refer to it as the \textit{band limit}. To  calculate numerically the integral in \Eqref{alm} over a finite coordinate grid, one requires a quadrature rule and knowledge of the SWSH over that grid. The quadrature rule presented in \cite{Huffenberger:2010hh} is based on a smooth non-invertible map where geometrically the poles are expanded as circles in $\mathbb{T}^2$.
Once a quadrature rule on equidistant points on $\mathbb{T}^2$ has been specified, we proceed to compute the SWSH which are written in terms of the so called \textit{Wigner d-matrices} \cite{Risbo:1996iy} by
\begin{equation*}\label{sYlm}
 _s Y_{lm} = (-1)^s \sqrt{\dfrac{2l+1}{4 \pi}} \s e^{\ii m \varphi} d^{l}_{m,-s}(\vartheta).
\end{equation*}
These matrices are easily calculated using recursion rules introduced by \cite{Trapani:2006he}. Adopting the notation $\Delta^{l}_{mn}  := d^{l}_{mn} ( \frac{\pi}{2} ) $, the Wigner d-matrices can be expressed as
\begin{equation}\label{Wigner}
d^{l}_{mn}(\vartheta) = \ii^{m-n} \sum \limits^{l}_{q=-l} \s \Delta^{l}_{qm} \s
e^{-\ii q \vartheta} \Delta^{l}_{qn},
\end{equation}
where  $n$ and $m$ take integer values that run from $-l$ to $l$. Later in \Sectionref{sec:computationWM}, we explain in detail how to compute the $\Delta^{l}_{nm}$ terms. The above expression allows to write the \textit{forward} and \textit{backward spin-weighted spherical harmonic transforms} respectively as
\begin{eqnarray}
_s a _{lm} &=& \ii^{s-m} \sqrt{\dfrac{2l+1}{4 \pi}} \s \sum\limits^{l}_{q=-l} \Delta^{l}_{qm} I_{qm} \Delta^{l}_{qs}, \label{eq:almHWT}\\ 
  f(\vartheta,\varphi) &=&\sum_{m,n} e^{\ii m\vartheta}e^{\ii n\varphi} J_{mn}, \label{eq:funcHWT}
\end{eqnarray}
where the matrices $I_{mn}$ and $J_{mn}$ are computed from the standard $2$-dimensional Fourier transforms (forward and backward, respectively) over $2\pi$-periodic extensions of the function $f(\vartheta,\varphi)$ into  $\mathbb{T}^2$.  In general, the complexity of the outlined algorithm is $\mathcal{O}(L^3)$.

\subsection{Axially symmetric spin-weighted transforms}\label{subsec:axialsymmetrictrans}

\subsubsection{The axially symmetric spin-weighted forward transform}
Let us begin by  pointing out that the previous algorithm can be decomposed into two main tasks; namely, computation of the $\Delta^{l}_{mn}$ terms and calculation of the $I_{mn}$ and $J_{mn}$ matrices by means of the $2$-dimensional forward and backward Fourier transforms, respectively, acting on some given function $f(\vartheta,\varphi)$. In what follows, we discuss in detail how to simplify these tasks for the case of axially symmetric functions, i.e., functions that only depend on the $\vartheta$ coordinate.

Let us consider a square integrable axially symmetric spin-weighted function
$f(\vartheta) \in L^2(\mathbb{S}^2)$ with spin-weight $s$. Due to the $\varphi$
dependence of the non-zero $m$ modes of SWSH, see \Eqref{sYlm}, we can write the
function $f(\vartheta)$ in terms of only ${}_s Y_{l0}(\vartheta,\varphi)$. Hence, the
forward spin-weighted spherical harmonic transform \Eqref{alm} can be written in
a simple form as
\begin{equation*}
_s a_{l} = \int \limits_{\mathbb{S}^2} \s f(\vartheta) \s { _s \overline Y_{l}}(\vartheta) \s \sin\vartheta\, d\vartheta \, d\varphi, 
\end{equation*}
where we have used the notation $_s a_{l}= {_sa_{l0}}$ and  $\s { _s  Y_{l}}(\vartheta) = \s { _s  Y_{l0}}(\vartheta,\varphi)$. Then, we rewrite  \Eqref{eq:almHWT} as
\begin{equation}\label{eq:almPartialATS}
_s a _{l} = \ii^{s} \sqrt{\dfrac{2l+1}{4 \pi}} \s \sum\limits^{l}_{n=-l} \Delta^{l}_{n0} I_{n} \Delta^{l}_{ns},
\end{equation}
with\footnote{The factor $2\pi$ comes from the trivial integral over $\varphi$.}
\begin{equation}\label{In}
I_{n} :=  2 \pi \int \limits^{ \pi}_{0} \s e^{-\ii n \vartheta} \s   f(\vartheta) \s \text{sin} \vartheta \s d\vartheta.
\end{equation}
Similarly  to what is done for HWTs in \cite{Huffenberger:2010hh}, the number of computations required to obtain the spectral coefficients ${}_s a_{l}$ can be reduced  by a factor of $2$ by using symmetries associated with the  $\Delta^l_{mn}$ quantities. In addition, we can introduce another reduction due the  fact that $\Delta^l_{n0} = 0$ for $l+n = \text{odd}$. This allows to reduce  the number of computations by a further factor of $2$. Therefore, we define \textit{the axially symmetric spin-weighted forward transform} (ASFT) as
\begin{equation}\label{eq:almAST}
{}_s a _{l} = \ii^{s} \sqrt{\dfrac{2l+1}{4 \pi}} \s \sum\limits^{l}_{n = l ( \text{mod}2 ) } \Delta^{l}_{n0} J_{n} \Delta^{l}_{ns} \hspace{0.1cm} \quad ( n +\!\!= 2 ),
\end{equation}
where $n$ is a positive integer that increases in steps of two and starts at $0$
or $1$ depending on whether $l$ is even or odd. The vector $J_{n}$ is  defined by
\begin{equation}\label{ec:Jn}
J_{n}:=
  \begin{cases}
    I_{n} & \mbox{ }n=0,\\
    I_{n}+(-1)^{s}I_{(-n)} & \mbox{ }  n>0.
  \end{cases}
\end{equation}
\noindent The evaluation of $I_{n}$ can be carried out  by extending  the function $f(\vartheta)$ to  $\mathbb{T}=\mathbb S^1$ as a $2\pi$-periodic function. This allows the implementation of the standard $1$-dimensional Fourier transform in contrast to the general case of HWTs which, due to the $\varphi$ dependence, requires a $2$-dimensional Fourier transform. Now, let us define the extended function on $\mathbb{T}$ as
\begin{equation}\label{ec:F_extended_on_T}
_s F(\vartheta) :=  \left \{ \begin{matrix} \hspace{-2.0cm}  f(\vartheta) & \text{ } \vartheta \leq \pi ,
\\  (-1)^s \s f( 2 \pi - \vartheta)  & \text{ } \vartheta > \pi . \end{matrix} \right .   
\end{equation}
Clearly, the vector $I_{n}$ remains unchanged because $_s F(\vartheta)$ agrees with $f(\vartheta)$ on the integration domain in \Eqref{In}.
\begin{figure}[t]
     \centering
    \includegraphics[scale=0.25]{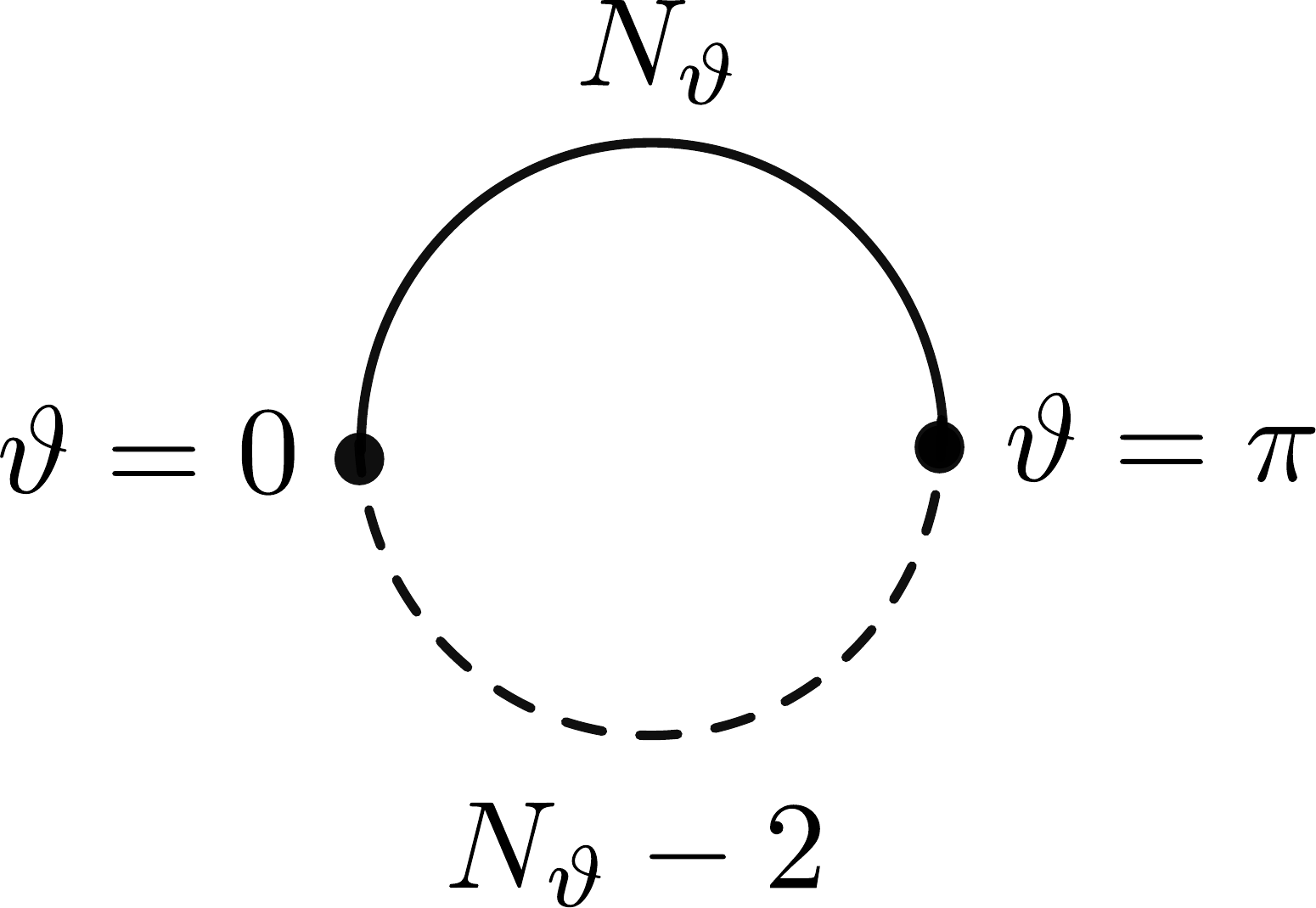}  
  \caption{}\label{sampling}
\end{figure}
The function $_s F(\vartheta)$ is chosen to be $2 \pi$ periodic, hence it can be
written as a $1$-dimensional Fourier sum. However, before doing so we need to
define the number of sampling points in $\mathbb{T}$. Let us consider
\Figref{sampling}. In this diagram, the upper part of the circumference
represents the number of samples $N_\vartheta$ taken for
$0 \le\vartheta \le \pi$, whereas the lower part shows the $N_\vartheta -2 $
samples for $\pi < \vartheta < 2 \pi $. Clearly, the subtraction by $2$ in the
lower half of the circumference comes from the extraction of the poles to avoid
oversampling. Therefore, to sample a function on $\mathbb{T}$ we proceed as
follows. If the desired number of samples for a function $f(\vartheta)$ on
$\mathbb{S}^2$ is $N_{\vartheta}$, then the number of samples for the extended
function $_s F(\vartheta)$ on $\mathbb{T}$ should be
$N'_\vartheta=2(N_\vartheta - 1 )$ and the spatial sampling interval will be
$\Delta \vartheta=2 \pi / N'_{\vartheta} $.  Therefore, the extended function
can be written as a $1$-dimensional Fourier sum by
\begin{equation*}
_s F(\vartheta) =\sum\limits^{N'_{\vartheta}/2}_{k=-N'_{\vartheta}/2+1} \s F_{k} e^{\ii k \vartheta}.
\end{equation*}
The substitution of this equation into \Eqref{In} yields
\begin{equation}\label{In2}
I_{n}= 2 \pi \sum\limits^{N'_{\vartheta}/2}_{k=-N'_{\vartheta}/2+1} \s F_{k} w(k-n),
\end{equation}
where $w(p)$ is a function $\mathbb{Z} \to \mathbb{R}$ defined  by 
\begin{equation*}
w(p)= \int \limits^{\pi}_{0} e^{\ii p \vartheta } \s \mbox{sin} \vartheta \s d\vartheta =
  \begin{cases}
     2/(1-p^2)  &  p  \mbox{ even} ,\\
     0      &  p \mbox{ odd, } p \not= \pm 1 , \\
    \pm \ii\pi /2 & p=\pm 1.
  \end{cases}
\end{equation*}
By comparison with \Eqref{In2} we note that the latter is proportional to a discrete convolution in the spectral space. Therefore, it can be evaluated as a multiplication in the real space such that $I_n$ is the $1$-dimensional forward Fourier transform of $2 \pi \: {}_sF \: w_r $ as follows
\begin{equation*}
  I_{n}=
  \frac{2\pi}{N'_\vartheta }
  \sum^{N'_\vartheta-1}_{q'=0}  
  \exp\left(-\ii n q\Delta\vartheta\right)\:
  {}_{s}F(q'\Delta\vartheta) \:
  w_r\left(q'\Delta\vartheta\right) ,
\end{equation*}
where $w_r(q' \Delta \vartheta)$ is the real-valued quadrature weight in $\mathbb{T}$ given by 
\begin{equation*}  
w_r(q' \Delta \vartheta) = \sum \limits^{N'_{\vartheta}/2
}_{p=-N'_{\vartheta}/2 +1} \s e^{-\ii pq' \Delta \vartheta } \s w(p) .
\end{equation*}
\noindent Finally, we want to emphasize that even though this way of sampling
functions on $\mathbb{T}$ allows to include the value of the extended function
at the poles, it yields an even number of modes in  spectral space. Hence, we
will not have the same number of positive and negative modes after application
of ASFT. Indeed, for the mode $I_{N'_{\vartheta}/2}$ (see \Eqref{ec:Jn}), the
vector $J_{L'}$ cannot be calculated since the term $I_{-N'_{\vartheta}/2}$ is
not given by the $1$-dimensional forward Fourier. We avoid this issue by
calculating the  set of $J_n$ terms  up to ${n=N'_{\vartheta}/2-1}$. Note that
setting $I_{N'_{\vartheta}/2}$ to zero does not constitute a loss of information
due to the exponential decay of the spectral coefficients of the Fourier
transform. In fact, this extra mode is in general numerically negligible and
hence, it will not affect the accuracy of the ASFT.

Now, in order to satisfy the Nyquist condition \cite{Fornberg:1998gv}, the relation between the number of sampling points in $\mathbb{T}$ and the band limit must satisfy the inequality
\begin{equation*}\label{eq:shannonrule}
2(N_\vartheta - 1 ) \ge ( 2 L + 1 ) + 1 ,
\end{equation*}
where the last  term on the right-hand side  comes from counting  the extra term without mirrored partner. As a result, the maximum value of the band limit for which the ASFT is exact is  
\begin{equation}\label{eq:optimalbandlimit}
L = N_\vartheta - 2 . 
\end{equation}

\subsubsection{The axially symmetric spin-weighted backward transform}\label{sec:algobwdXform}
This transform maps the spectral coefficients ${}_sa_{l}$  back to the corresponding axially symmetric function on $\mathbb{S}^2$. As the inverse transform does not contain integrals, issues of quadrature accuracy do not arise.
In a similar way as we  implemented  the  properties of the $3$-dimensional $\Delta^{l}_{nm}$ term to obtain \Eqref{eq:almAST}, we can write from \Eqref{eq:funcHWT} \textit{the axially symmetric spin-weighted backward transform} (ASBT) as 
\begin{equation*}
  f(\vartheta) = \sum\limits_{n=-N'_{\vartheta}/2 +1 }^{N'_{\vartheta}/2} e^{\ii n\vartheta} G_{n},
\end{equation*}
where the vector $G_n$ is given by
\begin{equation} \label{eq:bwdFSH}
G_{n}:=
  \begin{cases}
    0  & \text{if } n = N'_{\vartheta}/2 ,\\
    \ii^{s} \sum\limits_{l\equiv\text{mod}_{2}(n)}^{L} \sqrt{\frac{2l+1}{4\pi}} \hspace{0.1cm} \Delta^l_{n(-s)} \hspace{0.15cm} {}_sa_{l} \hspace{0.1cm} \Delta^l_{n0}\hspace{0.1cm}  & (l+\!\!=2)  .   
  \end{cases}
\end{equation}
Similar to \Eqref{eq:almAST},  $l$ increases in steps of two and starts at $l (
\text{mod}2 )$. We set $G_{N'_{\vartheta}/2} = 0$ because in the implementation
of the ASFT we chose $I_{N'_{\vartheta}/2} = 0$. The evaluation of
\Eqref{eq:bwdFSH} is carried out by a 1-dimensional inverse Fourier transform
that results in a function ${}_sF(\vartheta)$ sampled on $\mathbb{T}$. This function satisfies the symmetry properties in \Eqref{ec:F_extended_on_T} where  $f(\vartheta)$ represents the function ${}_sF(\vartheta)$ on $0\le \vartheta\le\pi$. Thus, ${}_sF(\vartheta)$ corresponds to the extension  of the  function $f(\vartheta)$ on $\mathbb{T}$.

\subsubsection{Computation of the \texorpdfstring{$3$-dimensional $\Delta^{l}_{nm}$}{3-dimensional Delta terms}}\label{sec:computationWM}
So far  the forward and backward spin-weighted spherical harmonic transforms have been simplified for axially symmetric functions by the implementation of a $1$-dimensional Fourier transform instead of a $2$-dimensional one as required in the algorithm HWT. In fact, we can simplify  the computation of the $\Delta^{l}_{nm}$ terms even further. This has a significant effect on the efficiency of both ASFT and ASBT, given that such a task takes around  half of the execution time in practical situations. We therefore devote this section to discuss this issue. 

\begin{figure}[t]
  \begin{minipage}{0.49\linewidth}
    \centering
     \includegraphics[width=\linewidth]{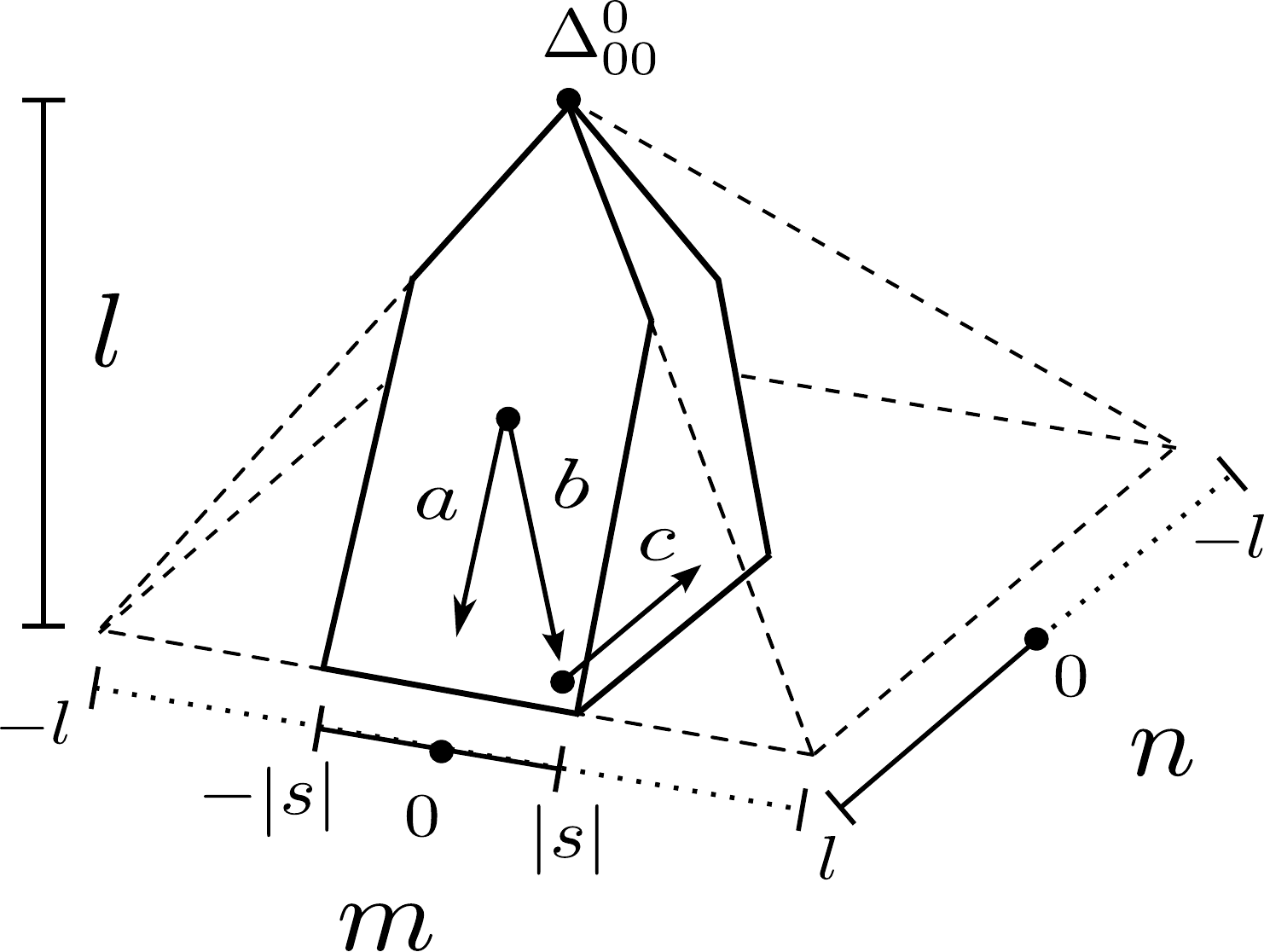}  
     \caption{} \label{piramide}    
  \end{minipage}
  \hfill
  \begin{minipage}{0.49\linewidth}
    \centering
    \vspace{0.6cm}
    \includegraphics[width=0.552\linewidth]{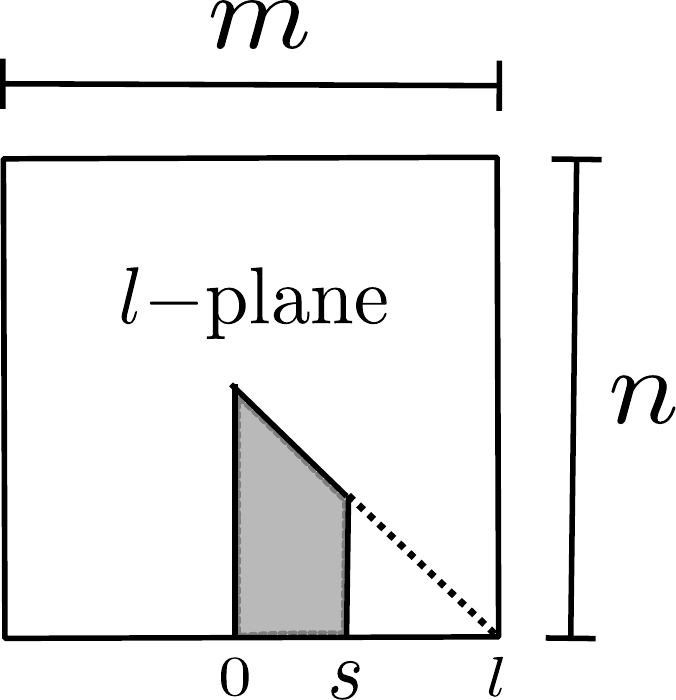}  
    \vspace{0.7cm}
     \caption{} \label{l-plane}
  \end{minipage}     
\end{figure}

Before we explain how the $\Delta^{l}_{nm}$ terms are computed, we bring up a relevant fact for both ASFT and ASBT. By examination of \Eqref{eq:almAST} and \Eqref{eq:bwdFSH}, we realize that we do not really need to calculate the complete set of $\Delta^{l}_{nm}$ terms\footnote{ $n$ and $m$ take integer values from $-l$ to $l$. } to perform the transform. Instead, we just need to compute up to the $\Delta^{l}_{ns}$ term, where $s$ is the spin-weight of the function that is supposed to be transformed. This yields a remarkable speed-up of the algorithm since in most cases $s\ll L$. Now, based on this, we proceed to compute the $\Delta^{l}_{ns}$ terms implementing the recursive algorithm introduced by Trapani and Navaza in \cite{Trapani:2006he}. The recursive relations are given by the following equations
\begin{eqnarray*}\label{eq:Deltasrecursion} 
(a)& &\Delta^{l}_{l0} \quad = \sqrt{\frac{2l-1}{2l}} \Delta^{l-1}_{(l-1)0}, \\
(b)& &\Delta^{l}_{lm} \quad = \sqrt{\frac{ l(2l-1)}{2(l+m)(l+m-1)}} \Delta^{l-1}_{(l-1)(m-1)}, \\ 
(c)& &\Delta^{l}_{nm} \quad = \frac{ 2 m }{  \sqrt{ (l-n)(l+n+1) }} \s \Delta^{l}_{(n+1)(m)} - \sqrt{ \frac{(l-n-1)(l+n+2)}{(l-n)(l+n+1)}} \s \Delta^{l}_{(n+2)(m)} ,
\end{eqnarray*}
where the  letters ``$(a)$'', ``$(b)$'' and ``$(c)$'' denote the sequence in which they should be used. We note that terms with a combination of indices outside of the correct range are set to $0$. One way to visualize the above algorithm is by means of the pyramidal representation of the $\Delta^{l}_{ns}$ terms in \Figref{piramide}. 
The volume of the complete pyramid represents the complete set of the
$\Delta^{l}_{nm}$ terms. Setting the top peak of the pyramid as
$\Delta^{0}_{00}=1$, we start moving down both in the vertical direction using
rule $(a)$, and in the diagonal direction by $(b)$. Thus, one can find the
$\Delta^{l}_{ns}$ terms in the right-hand side in the front face of the
pyramid. Then, using rule $(c)$ repeatedly, one can find the terms behind the
front face in order to calculate the right-hand side of the pyramid volume. If
we need to compute the full set of $\Delta^{l}_{nm}$ terms, we would need
to repeat this algorithm in order to obtain the complete right-hand side of the
pyramid volume. However, we just need to repeat step $(b)$ until we reach the
row corresponding to $l=|s|$ (for the given $l$-level) because we are just
interested in computing the first $\Delta^{l}_{ns}$ terms. Moreover, since only
the $\Delta^{l}_{ns}$ terms with positive values of $n$ are needed to compute
both ASFT and ASBT (see \Eqsref{eq:almAST} and (\ref{eq:bwdFSH})), we apply rule
$(c)$ until we reach the column $n=0$. The left-hand side of the pyramid volume
can be obtained by applying the mirror rule
$ \Delta^{l}_{n(-|s|)} = (-1)^{l-n} \Delta^{l}_{n|s|}$ (see
\cite{Trapani:2006he}). In \Figref{l-plane}, we display a schematic
representation of this, where the number of $\Delta^{l}_{ns}$ terms that have to
be computed are represented by the gray section. In this illustration we consider
the collection of the $\Delta^{l}_{ns}$ terms for each \textit{l-plane}. Note
that the gray section is not a rectangle since we can implement the symmetric
transposition rule $\Delta^{l}_{n|s|} = (-1)^{|s|-n} \Delta^{l}_{|s|n}$. In
short, we will require $\mathcal{O}(L^2)$ operations to compute the
$\Delta^{l}_{ns}$ terms needed for implementing both ASFT and ASBT, which will
allow us to pre-compute the $\Delta^{l}_{nm}$ terms for a low memory cost in
comparison with the general algorithm for HWTs\footnote{For $L = 1024$, the
  memory cost of AST is $\backsim 1$MB whereas for HWT is $\backsim 1$GB. }.

In conclusion, we have presented both the forward and backward spin-weighted
spherical harmonic transform for the axi-symmetric case by implementing
simplifications of the general algorithm HWTs in order to optimize them for
axially symmetric functions in \St. The first main simplification is the
replacement of the $2$-dimensional by a $1$-dimensional Fourier transform for
both the forward and backward transforms. This reduces the number of
computations to $\mathcal{O}( L \log_2 L)$. The second simplification
lies in the fact that both the forward and backward transforms do not need the
full set of $\Delta^{l}_{nm}$ terms in the axial case. Therefore, the resulting
algorithm requires $\mathcal{O}(L^2)$ operations for each transform. However, if
we precompute the Wigner coefficients $\Delta^l_{mn}$ then the transform is only
requires $\mathcal{O}(L \log_2L)$ operations.

These transform have been implemented in a Python 2.7 module\footnote{This module can be freely downloaded under the GNU General Public License (GPL) at  \url{http://gravity.otago.ac.nz/wiki/uploads/People/Axial_Spin_Weight_Functions.zip}}. Furthermore, the module allows to define objects that represent spin weighted functions for which an algebra can be defined. Hence, it can be seen not only as a set of functions, but as a Python environment for working with axi-symmetric SWSH.

\subsection{Choosing the optimal grid size}\label{sec:choosing_grid}
Because the axially symmetric transforms are based on the Fourier
transform, we expect that spectral coefficients decay exponentially to
zero when the band limit tends to infinity. Theoretically speaking, a
function is described in spectral space by an infinite number of
spectral coefficients. On the other hand, because of the machine
rounding error\footnote{In this paper, the terms ``machine rounding
  error'' and ``machine precision'' refer to the finite precision by
  which numbers can be represented in a computer. We always assume
  that this precision is of the order $10^{-15}$ which corresponds to
  standard ``double precision''.}, any sufficiently smooth function is
described by a finite set of spectral coefficients that contribute
numerically to the spectral decomposition. In other words, the
spectral coefficients with order lower than $10^{-15}$ are negligible
numerically, and thus are not necessary for an accurate description of
functions in the spectral space. Hereafter, we call the $l$-order of
the last mode above order $10^{-15}$ as the \textit{optimal band
  limit}. Consequently, in virtue of \Eqref{eq:optimalbandlimit} the
optimal band limit defines the \textit{optimal number of grid
  points}. Taking a larger number of grid points than the optimal one
will add unnecessary computations in the transform and consequently
the accuracy is reduced instead of enhanced. We refer to this as the
sampling error.  In our implementation we control this error by
keeping the number of grid points as close as possible to the optimal
case. To this end, we proceed as follows. Initially, we sample the
initial data in a large grid. In our case we have chosen
$N_{\theta} = 1025$. Then, we apply the ASFT to each function of the
initial data and identify the highest mode which is just above the
threshold $10^{-15}$. In other words, we identify the optimal band
limit for each function of the system.  From all these modes we set
the order of the highest mode as the optimal band limit for the
initial data. Henceforth we will refer to this as \textit{the global
  optimal band limit}. Using \Eqref{eq:optimalbandlimit} we obtain the
optimal number of grid points required to sample the functions of the
system. Finally, we begin the numerical solution of the system by
interpolating the initial data in the optimal grid.  Now, we discuss
how we keep the optimal grid size during the evolution. For each time
step, we check the last mode of each field in order to observe whether
they are smaller than some given tolerance. For this implementation
this has been set to $10^{-14}$. Then, if some of those modes do not
satisfy the mentioned condition, it implies that the number of grid
points is not enough for sampling some of the functions of the
system. Therefore, we need to interpolate all the functions to a
bigger grid.  We point out that the new grid should not differ too
much from the previous one because, as we mentioned before, it could
lead to too many unnecessary grid points and hence to larger
errors. In this implementation, we decided to increase the grid by
four points each time this is required. Using this small increment, we
expect to stay close enough to the optimal grid and as a consequence
keep a good accuracy.

To finalize this section we point out that due to non-linearities in our
evolution equations, some kind of filtering process is required in order to
avoid the so-called aliasing effect. For this we use the well known
2/3-rule. For details and justification  of this rule, see
\cite{Fornberg:1998gv} and references therein.

\section{Numerical application}\label{sec:Application_test}

\subsection{Smooth Gowdy symmetry generalized Taub-NUT solutions}\label{sec:classtaubNut}
It is well known that solutions of Einstein’s field equations are
uniquely determined (up to isometries and questions of extendibility)
by the Cauchy data on a Cauchy surface. However, there exist cases for
which the uniquely determined maximal globally hyperbolic development
\cite{ChoquetBruhat:1969cl} of the data can be extended in several
inequivalent ways. These extensions are not globally hyperbolic and
hence there are \textit{Cauchy horizons} whose topology and smoothness
may in general be complicated. Furthermore, there can exist closed
causal curves in the extended regions which violate basic causality
conditions. A well known example of this sort of solutions is the Taub
solution \cite{Taub:1951vk}, which is a two-parametric family of
spatially homogeneous cosmological models with spatial topology
\Sth. Extensions through the Cauchy horizons are known as Taub-NUT
solutions \cite{Newman:1963up}.

As generalizations of the Taub(-NUT) solutions, we consider now the
class of \textit{smooth Gowdy-symmetric generalized Taub-NUT
  solutions} introduced in \cite{Beyer:2011uz}
motivated by early work by Moncrief \cite{Moncrief:1984js}. These are
Gowdy-symmetric globally hyperbolic solutions of Einstein's vacuum
field equation with zero cosmological constant and spatial topology
\Sth which have a past Cauchy horizon with topology \Sth ruled by
closed generators. To cover the maximal global hyperbolic
developments, the class is written in terms of the ``areal'' time
function $t \in (0, \pi)$ \cite{Chrusciel:1990ti} and the same Euler
coordinates as in \Sectionref{subsec:s3topology} for the spatial
part. In these coordinates the metrics take the form
\begin{equation}\label{class_Gowdy_metric}
g = e^M ( -\df t^2 + \df \theta^2 ) + R_0 \left(  \sin^2t \text{ } e^u (\df \rho_1 + Q \df \rho_2 )^2 + \sin^2 \theta \text{ }  e^{-u}  \df \rho^{2}_2  \right),   
\end{equation}
with a positive constant $R_0$ and smooth functions $u$, $Q$ and $M$
that depend only on $t$ and $\theta$. A large class of such solutions
of the Einstein vacuum equations were constructed in
\cite{Beyer:2011uz} as an application of the Fuchsian method 
\cite{Ames:2012vz}.

\subsection{A family of exact solutions}
\label{sec:exactsol}
In a subsequent paper \cite{Beyer:2014vw}, the same authors introduced
a three-parametric family of explicit smooth Gowdy-symmetric
generalized Taub-NUT solutions as an application of soliton
methods. For this family of exact
solutions, the components of the metric \Eqref{class_Gowdy_metric} are
given by
\begin{eqnarray*}   
& & e^M  = \dfrac{R_0}{64 c^{3}_{1} } \left( U^2 + V^2 \right), 
\quad e^u = \dfrac{R_0}{64 c^{2}_{1} } \dfrac{ U e^{-M} }{ 1 + y },\\
& & Q  =  x + \dfrac{ c_3 }{8} \left( 1- x^2 \right) \left( 7 + 4y + y^2 + \dfrac{(1-y)V^{2}}{4 c^{2}_{1} U } \right),
\end{eqnarray*}
where 
\begin{equation*}
U= c^{2}_{3} \left(  1-x^2 \right) \left( 1-y \right)^3 + 4 c^{2}_{1}
(1+y), \quad 
V= 4 c_{1} \left(  1-y \right) \left(  1 - c_3  x(2+y) \right),  
\end{equation*}
with $x=\cos \vartheta $, $y=\cos t $. Here $c_1$ and $c_3$ are real
constants that, together with $R_0$, define particular
solutions. We point out that this family of  solutions  contains the
spatially homogeneous  Taub solutions as the special case 
given by 
\begin{equation*}
c_1 = \dfrac{1}{l}\left( \sqrt{l^2 +m^2 } + m \right), \quad 
c_3=0,\quad
R_0 = 2 l \sqrt{l^2 +m^2}, 
\end{equation*}
with free parameters $l>0$ and $m \in\R$. Inhomogeneous  solutions  are obtained by choosing any non-zero value for $c_3$ (see \cite{Beyer:2014vw} for details).

In the following we now perform the Geroch reduction described
in \Sectionsref{sec:gerochReduction} and \ref{subsec:s3topology} for these exact
solutions. As a consequence
of Gowdy symmetry, the vector field $\partial_{\varphi}$ is a
smooth Killing field of the $2+1$ metric $h_{ab}$. Consequently, all
the metric components of any Gowdy symmetric metric represented in
these two coordinates are axial symmetric in the sense defined
in \Sectionref{sec:numerical_implementation} and hence the axial
symmetric transform introduced
in \Sectionref{subsec:axialsymmetrictrans} is the natural choice for
our numerical implementation discussed below. Before we discuss this
in detail, we list the resulting formulas
\begin{eqnarray}
               \psi &=& R_0 \sin^2t \text{ }  e^u , \label{eq:psi} \\
\partial_t  \omega &=& - R_0 \text{ } \dfrac{\sin^3 t }{ \text{sin} \vartheta } \text{ } e^{2u} \partial_{\vartheta} Q, \label{eq:omega_t} \\
\partial_{\vartheta}  \omega &=& - R_0  \text{ } \dfrac{\sin^3 t }{\text{sin} \vartheta } \text{ } e^{2u} \partial_{t} Q, \label{eq:omega_theta} \\
h &=& \psi \left(e^{M} ( -\df t^2 + \df \vartheta^2 ) + R_0  \sin^2 \vartheta \text{ } e^{-u}  \df \rho^{2}_{2} \right),
\end{eqnarray}
where $\psi$ and $\omega$ are the norm and twist associated with
$\partial_{\rho_1}$ and $h_{ab}$ the $2+1$ metric. Next, as described
in \Sectionref{Sec:evol_equations}, we write the metric in terms of the
frame $(T,m,\overline{m})$ which yields
\begin{eqnarray}
\lambda &=& R_0  \sin^2 t \text{ } e^{M+u}, \label{eq:lambda} \\
\beta   &=& 0 , \label{eq:beta} \\
\delta  &=& R_0  \sin^2 t \text{ }   \left( e^{M+u} +  R_0  \right) / 2 , \label{eq:delta} \\
\phi    &=& R_0  \sin^2 t \text{ }  \left( e^{M+u} -  R_0  \right) / 2 . \label{eq:phi}
\end{eqnarray}
Henceforth, we  refer to $h_{ab}$ as the $2+1$ smooth Gowdy symmetry
generalized Taub-NUT metric. 

We notice that the quantities $\mathring{\Gamma}_{\mu}$ associated
with $h_{ab}$ are calculated from \Eqref{eq:gammas} by first computing the
contracted Christoffel symbols $\Gamma_\mu$ of $h_{ab}$ and then by calculating
$\breve{\Gamma}_\mu$ from \Eqref{eq:gammas2} and the background metric
\Eqref{eq:background_metric}. The results are 
\begin{eqnarray}
\label{eq:f0exact}
\mathring{\Gamma}_{0}&=& -\cot t , \\
\label{eq:f2exact}
\mathring{\Gamma}_{1}&=& \mathring{\Gamma}_{2}= \sqrt{2} \text{ }  c^{2}_{3} \text{ } \csc^2 t \text{ }
         \sin^8 \frac{t}{2} \text{ } \sin 2\vartheta.
\end{eqnarray}
Here, and in all of what follows, we choose $c_1 = 1$,
$R_0 = 2$, and only vary $c_3$.

For the following it is also convenient to list  the values of the
metric functions at the time $t=\pi/2$ which we shall use as the
initial data for our numerical evolutions. Notice that we cannot use
$t=0$ or $t=\pi$ as initial times because the data are singular there.
Thus, evaluating \Eqsref{eq:lambda}--\eqref{eq:phi} and time derivatives  at $t=\pi/2$, we obtain\footnote{We have used $\partial_t g_0$ to denote the temporal partial derivative of any function $g$ evaluated at the initial time.}
\begin{align}\label{initial_data_metric}
 \lambda_0 &=  -4- c_3^2 \sin^2 \vartheta,         & \partial_t \lambda_0 &= -4 c_3^2 \sin^2 \vartheta , \\
 \phi_0    &=  \frac{c_3^2}{2}  \sin^2 \vartheta,  & \partial_t \phi_0    &= c_3^2 \sin^2 \vartheta ,\\
 \delta_0  &=  4+\frac{c_3^2}{2}\sin^2 \vartheta,  & \partial_t \delta_0  &= c_3^2 \sin^2 \vartheta,\\
 \beta_0   &=  0 ,                                 & \partial_t \beta_0   &= 0 .
\end{align}
From \Eqref{eq:psi} and  its time derivative we obtain the initial
values for $\psi_0$ and $\partial_t \psi_0$ respectively. Finally, by
integrating \Eqref{eq:omega_theta} with respect to $\vartheta$ and
setting the irrelevant integration constant to zero we obtain
$\omega_0$. By considering \Eqref{eq:omega_t} we obtain  $\partial_t \omega_0$. The explicit form of these functions is 
\begin{eqnarray}
\omega_0 &=& \dfrac{-128 (-8+16 c_3 \cos\vartheta)}{256+288 c_3^2+3 c_3^4-512 c_3 \cos\vartheta-4 c_3^2 \left(-56+c_3^2\right) \cos2\vartheta+c_3^4 \cos 4\vartheta } \:, \\  
\psi_0 &=& \dfrac{8 \left(1+\frac{1}{4} c_3^2 \sin^2\vartheta\right)}{(1-2 c_3 \cos\vartheta)^2+\left(1+\frac{1}{4} c_3^2 \sin^2\vartheta \right)^2}\:,\\  
\partial_t \omega_0 &=& 128 \big(64+64 c_3^2 \cos^2\vartheta-64 c_3^3 \cos^3\vartheta-4 c_3^4 \sin^4\vartheta \\
                    & &  + c_3 \cos\vartheta \left(-128+8 c_3^2 \sin^2\vartheta+9 c_3^4 \sin^4\vartheta \right)\big)/ \: B \: ,\nonumber\\
\label{initial_data_normandtiwst}
\partial_t \psi_0 &=& - 64 c_3 \big(128 c_3 \cos^2\vartheta-32 c_3 \sin^2\vartheta+4 c_3^3 \sin^4\vartheta+c_3^5 \sin^6\vartheta \\
                  & & +16 \cos\vartheta \left(-12+5 c_3^2 \sin^2\vartheta \right)-24 c_3^3 \sin^2 2\vartheta \big) / \: B \:,\nonumber
\end{eqnarray}
where
\begin{equation*}
B= \left(32-64 c_3 \cos\vartheta+64 c_3^2 \cos^2\vartheta+8 c_3^2 \sin^2\vartheta+c_3^4 \sin^4\vartheta \right)^2 \: .
\end{equation*}

\subsection{Numerical error sources}
\label{sec:numerrorsources}
The purpose of the following subsections is to describe the numerical evolution
of the equations \eqref{eq:geroch_einstein_equations_final} for the
just discussed initial data. We shall do this for two sets
of gauge source functions. Before we go into the details
in \Sectionsref{sec:numareal} and \ref{sec:numharmonic}, however, let
us discuss possible numerical error sources which we shall refer to in our
discussion of our numerical results below.

Clearly the time and spatial discretization gives rise to numerical
errors. In general it is expected
that time discretization errors are larger than spatial ones thanks to
the rapid (exponential) convergence of the latter. In order to
investigate the presumably more significant time discretization errors
we shall use two different time discretization schemes, the
(non-adaptive) 4th-order Runge-Kutta scheme and the (adaptive)
Runge-Kutta-Fehlberg (RKF) scheme. See \cite{Durran:2010} for details
about adaptive Runge-Kutta methods. Spatial discretizations shall
always be based on our adaptive framework discussed
in \Sectionref{sec:choosing_grid}. For runs using the adaptive RKF
scheme, we can therefore expect that all discretization errors can be made
sufficiently small by choosing suitable tolerance parameters.

In our numerical experiments we identify further error sources
which turn out to be particularly severe.  Recall from \Sectionref{sec:choosing_grid} that we choose the same band limit
for \textit{all} unknowns. However, in most of our practical examples,
\textit{only few} of the unknowns actually require high spatial
resolutions. As a consequence, many unknowns are \textit{oversampled},
which is not only inefficient numerically, but also generates undesired
numerical noise.
The origin of this noise is that the ``unnecessary'' modes associated
with too large band limits are in general not zero numerically. In
fact, while they are typically of the order of the machine precision
initially, they may grow during the evolution in particular due nonlinear
coupling of modes. Typically, the larger the difference between the optimal
band limit for any particular unknown and the global band limit is, the larger is
this noise. This error is difficult to control in practice and it is
quite common that once this noise has started to grow during the
evolution it continues to grow increasingly fast. We measure this
error by looking at the evolution of the highest modes of certain
representative unknowns during the evolution.  The only conceivable
cure of this problem would be to work with higher machine precisions,
which would, however, significantly slow down the numerical runs. Our
numerical infrastructure is completely based on ``double
precision''. We have not attempted to work with higher machine
precisions such as ``quad precision'' yet. Further comments on this in
the context of a different numerical infrastructure can be found, for
example, in \cite{Beyer:2009vw}.

Another severe, but not fully independent numerical error is associated with the violation of the constraints. Recall that due to \Eqref{subsidiarysystem}, the condition
$\mathcal D_\mu\equiv 0$ is identically satisfied during the evolution
if (i) the evolution equations hold exactly and (ii) the constraints
\Eqsref{eq:constraint1} and \eqref{eq:constraint2} are satisfied
initially. For our numerical calculations, however, both of these
conditions are violated. Let us for the sake of this argument imagine
that the constraints are violated at the initial time, but that the
evolution equations hold identically (i.e., we pretend that the
numerical evolutions are done with an infinite resolution in space and
time and with infinite machine precision).  Then
\Eqref{subsidiarysystem} describes the (exact) evolution of the in
general non-zero constraint violation quantities $\mathcal D_\mu$.
Since the initial data for these quantities are now assumed to be
non-zero, their evolution is in general also
non-zero. Depending on the particular properties of the evolution
system and hence of \Eqref{subsidiarysystem}, these quantities may in
fact grow rapidly during the evolution. If this is the case, the
constraint violation error can become large very quickly even if it is
small at the initial time, and the resulting numerical solutions of
Einstein's equations therefore become useless quickly. This situation
cannot be improved by increasing the numerical resolution. In fact,
this error is a consequence of the structure of the \textit{continuum}
evolution equations.  Various ways to reconcile this problem have been
proposed in the literature. One of the most promising ideas
\cite{Brodbeck:1999en,Gundlach05,Beyer:2013uq} is to introduce
constraint damping terms, i.e., to add terms to the evolution
equations (i) which are proportional to the constraint violation
quantities (hence the solutions of the evolution equations for the
actual case of interest $\mathcal D_\mu\equiv 0$ are unchanged) and
(ii) which, however, turn the surface $\mathcal D_\mu\equiv 0$ into a
future attractor for \Eqref{subsidiarysystem}. This technique has
proved to be quite useful to produce stable calculations for
asymptotically flat spacetimes (see for instance
\cite{Preto2005:Evolution-of-Binary-Black-Hole,Lindblom:2004ci,Holst:2004ep,Boyle:2007kb}). The
analytic derivation of suitable constraint damping terms is in general difficult and is usually done based on
approximations which may only hold in certain regimes of the
evolution (see e.g., \cite{Frauendiener04}). In this paper here we work
without constraint damping terms. Nevertheless, we remark that thanks
to the close relationship of our
formulation of Einstein's equations with the ones used in the above
references, similar choices of constraint damping terms are expected
to be useful in reducing the constraint violation
errors in our applications. Indeed we have already gathered some promising 
experience with constraint damping terms of the type used in
\cite{Preto2005:Evolution-of-Binary-Black-Hole} which we shall report
on in a future article.

\subsection{Numerical evolutions in areal gauge}
\label{sec:numareal}
In this section now, we shall fix the gauge freedom for the evolution
equations by identifying the gauge source functions $f_\mu$ with the
contracted Christoffel symbols 
given by \Eqsref{eq:f0exact} -- \eqref{eq:f2exact}; we recall that we have implicitly always assumed  \Eqref{eq:background_metric} as the reference metric and continue to do so.  As common in the
literature, we refer to this coordinate gauge as \textit{areal gauge}.
We shall then evolve the evolution equations
\eqref{eq:geroch_einstein_equations_final} for the initial data given
by \Eqsref{initial_data_metric} -- \eqref{initial_data_normandtiwst}
at $t=\pi/2$ using these gauge source functions.  The resulting
numerical solutions are given in the same coordinates as the exact
solution and direct comparisons between the exact and the numerical
solutions can be performed conveniently by considering the error quantity
\begin{equation*}\label{eq:norm}
E(t):= \max_{\mu,\nu}  \: \| \: h^{(e)}_{\mu\nu}(t,\vartheta) - h^{(n)}_{\mu\nu}(t,\vartheta) \:  \|_{L^2(\St)}  \; ,
\end{equation*}
where $h^{(e)}_{\mu\nu}(t,\vartheta)$ represents a frame component of the
\textit{exact} metric for any given $t$, whereas $h^{(n)}_{\mu\nu}(t,\vartheta)$
represents the numerical value.  The norm $\| \cdot \|_{L^2(\St)}$ is
approximated numerically by the discrete $\ell^2$-norm of the grid function
vector.  Notice that the same spacetimes in the same coordinates have been
constructed numerically with different methods in \cite{Hennig:2013eu}. However, in contrast to our discussion here,
some of Einstein's equations turn out to be formally singular in the ``interior'' of the
Gowdy square  in the formulation
used there and hence are ignored to avoid serious numerical problems.

\begin{figure}[t]
    \centering
    \includegraphics[width=0.5\linewidth]{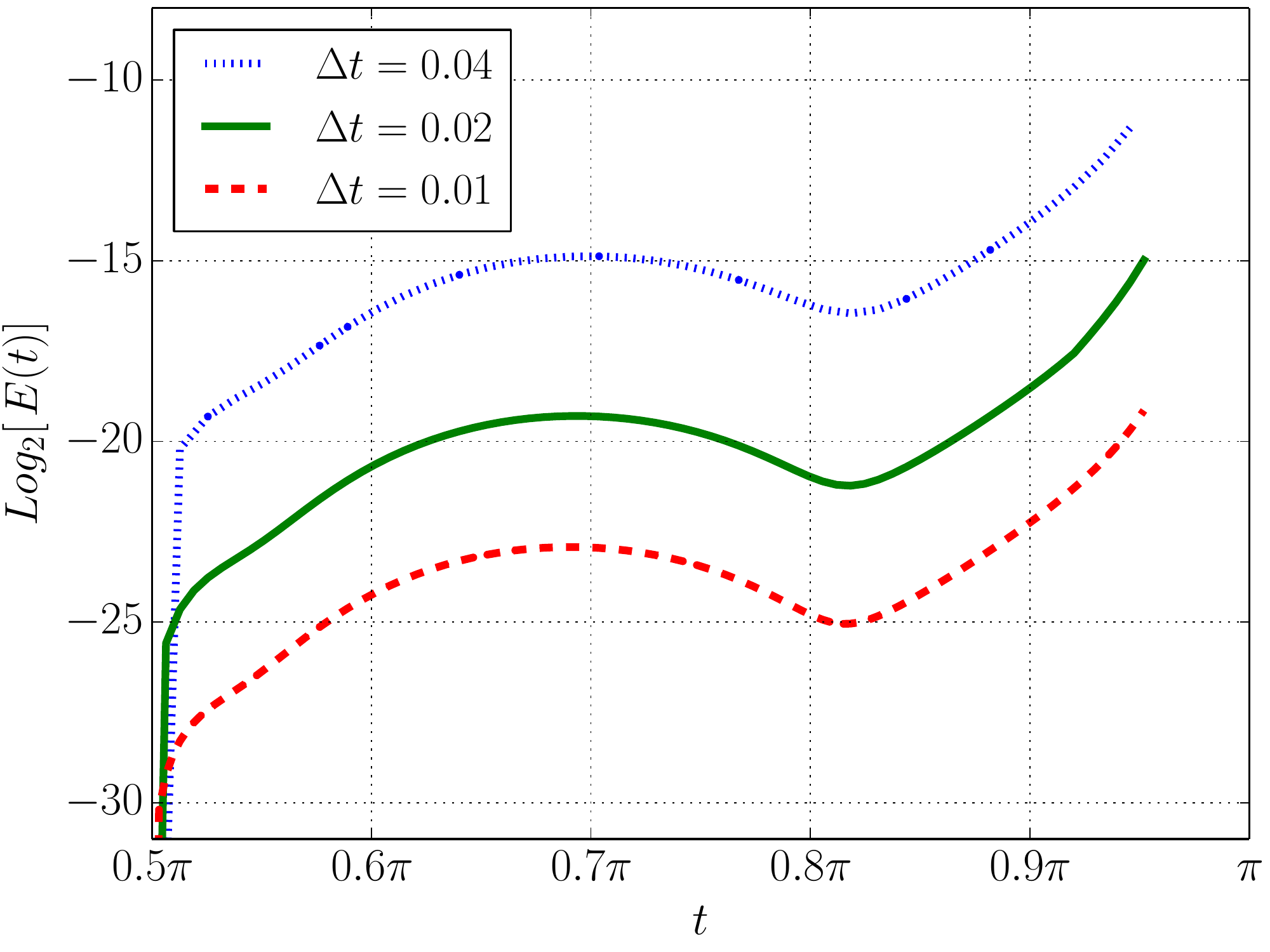}
    \caption{Convergence test, $c_3 = 0.2$.}
   \label{Convergencetest}
\end{figure}

As a first test for our numerical implementation we present a
convergence test in \Figref{Convergencetest} for $c_3 = 0.2$.  The
evolution is carried out with the (non-adaptive) 4th-order Runge-Kutta
scheme. The figure shows the expected convergence rate demonstrating
that the time discretization error is dominant here. This is not
surprising since at each $t$ all the metric components are very smooth
functions that can be resolved on the grid with high accuracy so long
as $t$ does not get too close to $t=\pi$. The oversampling and
constraint violation errors discussed in the previous subsection are
small during this early phase of the evolution.

\begin{figure}[t]
\begin{minipage}{0.49\linewidth}
    \centering
    \includegraphics[width=\linewidth]{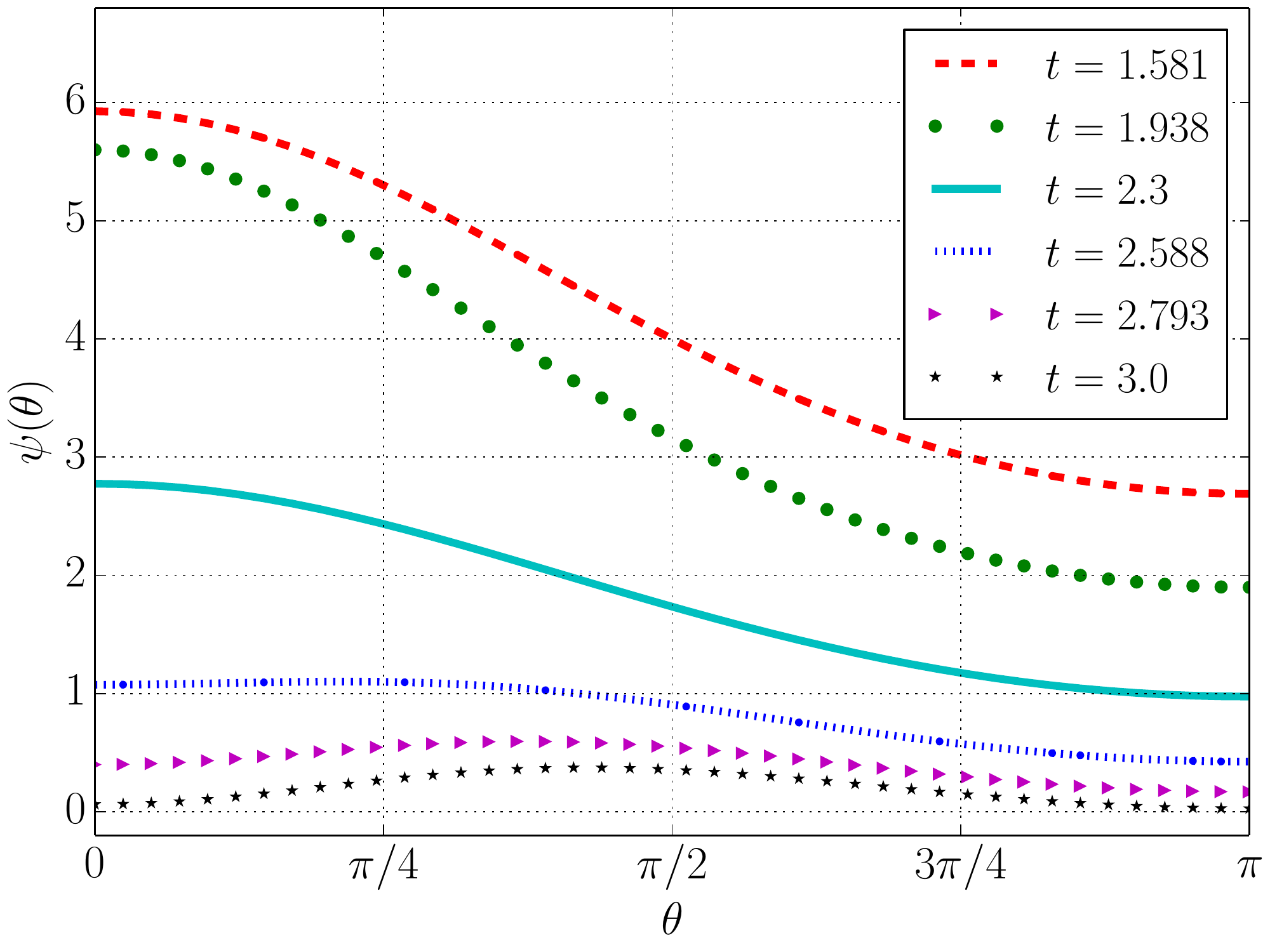}
    \caption{Norm of the Killing vector, $c_3 = 0.2$.}
    \label{KillingNorm}
  \end{minipage}
  \hfill
  \begin{minipage}{0.49\linewidth}
    \centering
    \includegraphics[width=\linewidth]{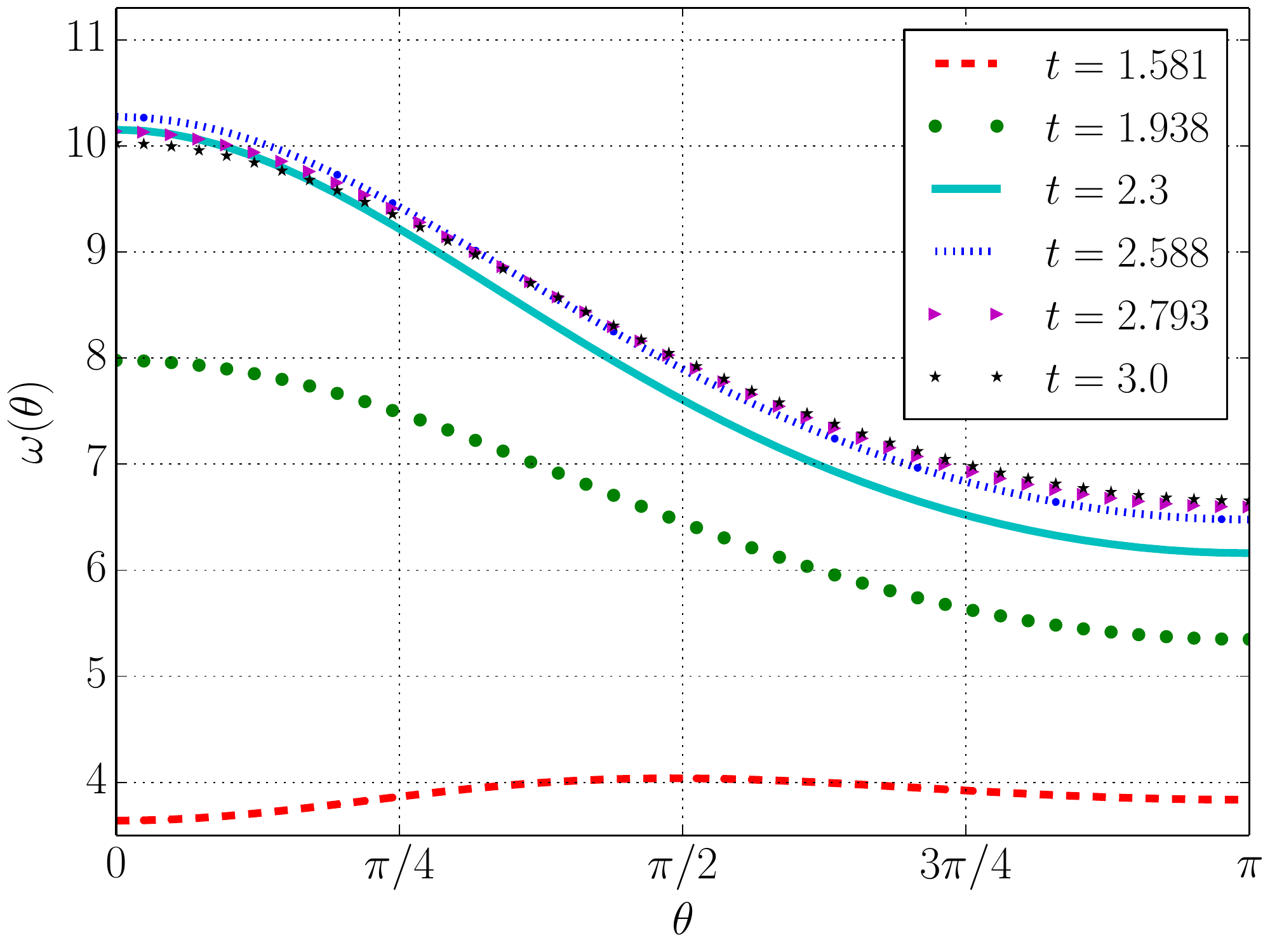}
    \caption{Twist of the Killing vector, $c_3 = 0.2$.}
    \label{Twist}
  \end{minipage}
\end{figure}

\begin{figure}[t]  
    \begin{minipage}{0.49\linewidth}
    \centering
    \includegraphics[width=\linewidth]{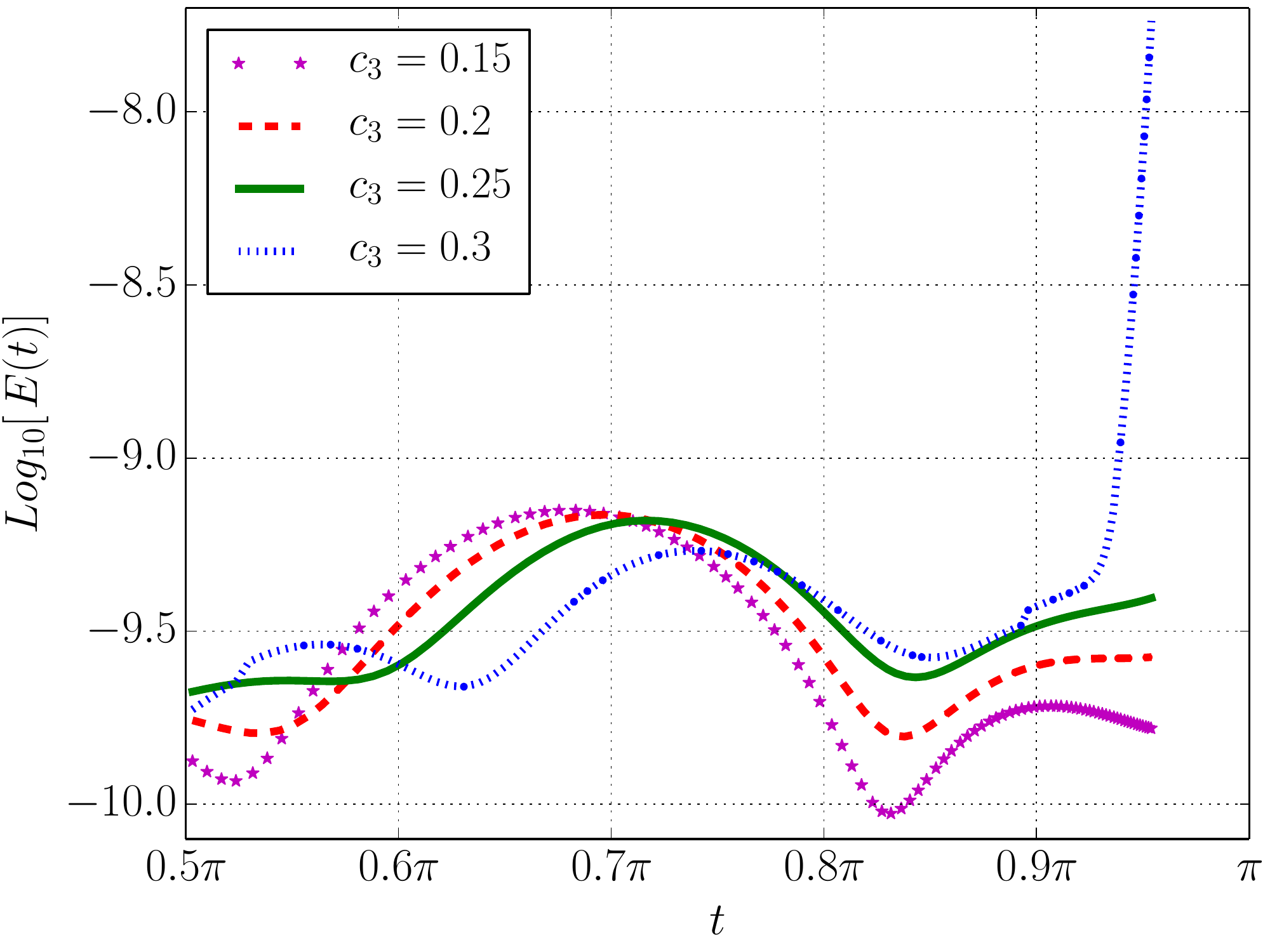}
    \caption{Error propagation for various values of $c_3$ and Tol = $10^{-8}$.}\label{f4N} 
    \end{minipage}
    \hfill
    \begin{minipage}{0.49\linewidth}
    \centering
    \includegraphics[width=\linewidth]{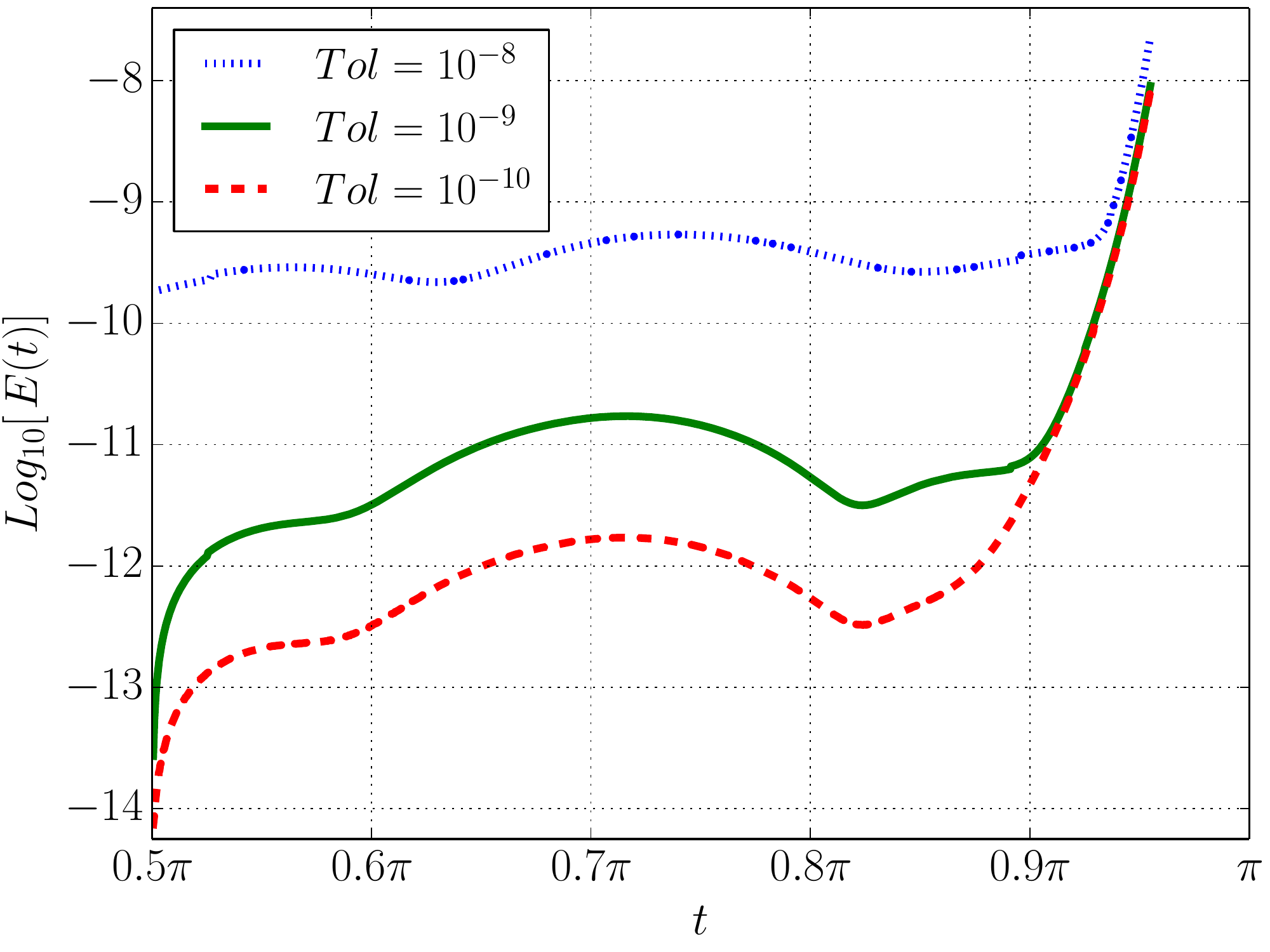}
    \caption{Error propagation for various values of $Tol$ and $c_3 = 0.3$.}\label{f5N} 
  \end{minipage}
\end{figure}

Next, we replace the non-adaptive 4th-order Runge-Kutta scheme by the
adaptive RKF method. In \Figref{KillingNorm} and \Figref{Twist} we
show the numerical evolutions of the geometric quantities $\psi$ and
$\omega$ for $c_3=0.2$.
The numerical error in these calculations are shown in \Figref{f4N}
for different values of $c_3$. The error tolerance \emph{Tol} of the RKF
method is chosen to be $10^{-8}$. This figure suggests that the
numerical errors here remain bounded for a long time. The larger
$c_3$ is, however, and hence the more inhomogeneous the solution is,
the more rapidly the numerical errors grow close to $t=\pi$ as
expected. \Figref{f5N} indicates that the behavior close to $t=\pi$
cannot be improved by decreasing the value of \emph{Tol}. This suggests
that close to $t=\pi$ the numerical errors are not dominated anymore
by the time discretization error, but that one of the other error
sources discussed in \Sectionref{sec:numerrorsources} takes over.
Our experience suggests that in fact both the oversampling error and
the constraint violation error are significant at late times, in
particular, for larger values of $c_3$. As a consequence of
\Eqsref{initial_data_metric} -- \eqref{initial_data_normandtiwst}, the
required band limits for the metric components and their time
derivatives are small, but the required band limits to resolve
$\psi_0$, $\omega_0$ and their time derivatives are relatively
large. This discrepancy, which we associate with the oversampling
error, is in fact larger the larger $c_3$ is. As already mentioned before, the noise generated by oversampling indeed grows during the
evolution. 

\begin{figure}[t] 
    \centering
    \includegraphics[width=0.5\linewidth]{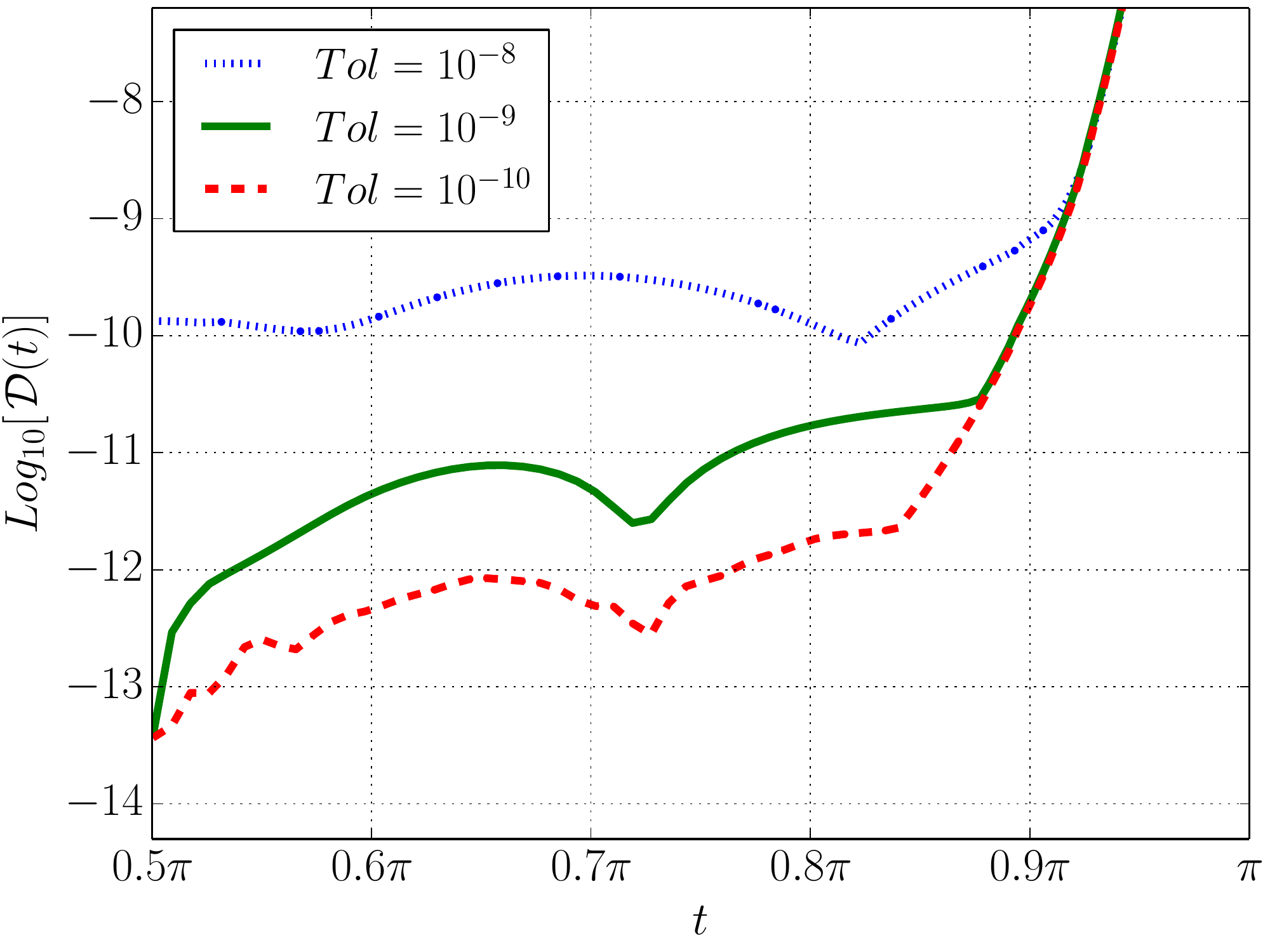}
    \caption{Constraints propagation for various values of \textit{Tol} and $c_3 = 0.3$.}\label{f6N}
\end{figure}
In order to measure the constraint violation error, we define the quantity
\begin{equation*}\label{eq:constraint_propagation}
\mathcal{D}(t) := \max_{\mu}  \: \| \:  f_{\mu}(t,\vartheta) - \mathring{\Gamma}_{\mu}(t,\vartheta) \:  \|_{L^2(\St)}.
\end{equation*}
In \Figref{f6N}, we show the evolution of this quantity for $c_3=0.3$.
At late times, the curves look very similar to the ones of  $E(t)$ in \Figref{f5N}. This
suggests that the constraint violation error contributes significantly
to the total numerical error.

\subsection{Numerical evolutions in wave map gauge}
\label{sec:numharmonic}

In this section we describe numerical computations for the same
spacetimes as before, but using a different coordinate gauge. To this end,
we want to choose the same initial data as before, but work with
different gauge source functions. Both the gauge constraint
\Eqref{eq:constraint1} and the physical constraints
\Eqref{eq:constraint2} clearly have to be satisfied at the initial
time. Since we do not want to resolve these complicated nonlinear
PDEs, our strategy is to use exactly the same initial data for the
values of the metric components and their first time derivative
values, and also exactly the same initial values of the gauge source
functions as before. In order to implement a different coordinate gauge,
we then apply the following ``gauge driver condition'' during the
evolution whose purpose is to rapidly drive the gauge source functions
from their \emph{initial values} fixed by the gauge constraint towards
the \emph{target gauge source functions} $\hat f_\mu$:
\begin{equation}
  \label{eq:gaugedriver}
  f_\mu= (\mathring{\Gamma}_{\mu} |_{t_0}-\hat f_\mu) e^{-q (t-t_0)}+\hat f_\mu.
\end{equation}
Here the parameter $q$ controls how rapidly the gauge is driven towards
the target. The quantities $\mathring{\Gamma}_{\mu} |_{t_0}$ are calculated from
the initial data and are understood as functions of the spatial coordinates
only.  Notice that different gauge drivers for the generalized wave
representation of Einstein's equations were considered in
\cite{Lindblom:2009in}. \Eqref{eq:gaugedriver}
guarantees that the gauge constraint is satisfied at the initial time. As
discussed at the end of \Sectionref{sec:Hyperbolicreduction}, the physical
constraints, even though they pose highly non-trivial restrictions on the choice
of the initial data because they are essentially linear combinations of the
well-known Hamiltonian and momentum constraints, turn out to \textit{not}
be restrictions on the gauge source functions. Hence it is not necessary to
introduce terms in \Eqref{eq:gaugedriver} which account for the first time
derivative of $\mathring{\Gamma}_{\mu}$ at $t=t_0$.

We apply this idea to calculate the same spacetimes as before, but now we choose
the \emph{wave gauge} as the target gauge, which is defined by the condition
$\hat f_{\mu}=0$. For our numerical tests we choose $q=10$ in
\Eqref{eq:gaugedriver}. Before we present our numerical results we notice
that it is straightforward to derive the formula
\begin{equation}\label{time_trans}
  t_{(w)}=\dfrac{\pi}{2}+ \dfrac{1}{2} \log \left( \dfrac{1-\cos t}{1+\cos t} \right),
\end{equation}
which for our spacetimes relates the time coordinate $t$ in areal gauge (used in
\Sectionref{sec:numareal}) and the time coordinate $t_{(w)}$ in wave map gauge. This
formula holds identically even though \Eqref{eq:gaugedriver} is strictly
speaking not the exact wave map gauge. However as a consequence of
$\mathring{\Gamma}_{0} |_{t_0=\pi/2}=0$ which follows from \Eqref{eq:f0exact},
the target gauge source function $\hat f_0=0$ agrees identically with $f_0=0$.
\Eqref{time_trans} is then obtained by solving the exactly homogeneous wave
map equation for the wave map time coordinate function with appropriate initial
conditions.  \Eqref{time_trans} allows us to make direct comparisons between our
results here and the results in the previous section.  In particular, it reveals
that the wave time slices $t_{(w)}=const$ are the same as the areal time
slices $t=const$ (for different constants), and, the ``singularities'' at
$t=0,\pi$ are shifted to infinity, in particular, $t_{(w)} \to \infty $ for
$t \to \pi$.  We point out however that it is not possible to derive a formula
which relates the spatial coordinates in both gauges. This is true even if $q$
in \Eqref{eq:gaugedriver} was so large that we could consider our gauge as
the exact wave map gauge. This is a consequence of the fact that the homogeneity of the wave
equations for the spatial wave map coordinates is destroyed by terms given by
the reference metric \Eqref{eq:background_metric}. In fact we shall demonstrate below
that the spatial coordinates on each time slice are different in areal and
wave map coordinates.

\begin{figure}[t]
  \begin{minipage}{0.49\linewidth}
    \centering
    \includegraphics[width=\linewidth]{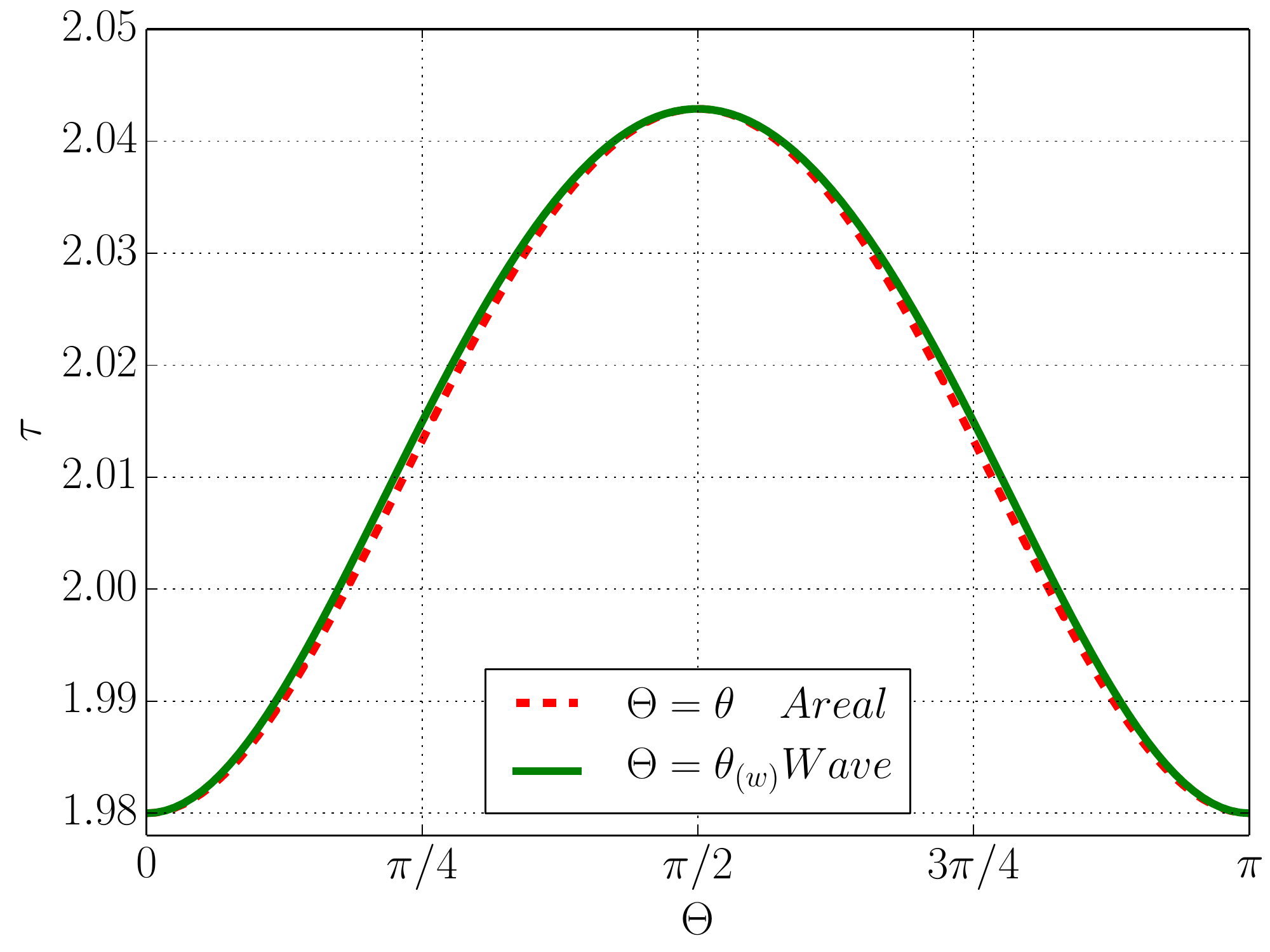} 
    \caption{Proper time comparison.}\label{f9N} 
  \end{minipage}
  \hfill
  \begin{minipage}{0.49\linewidth}
    \centering
    \includegraphics[width=\linewidth]{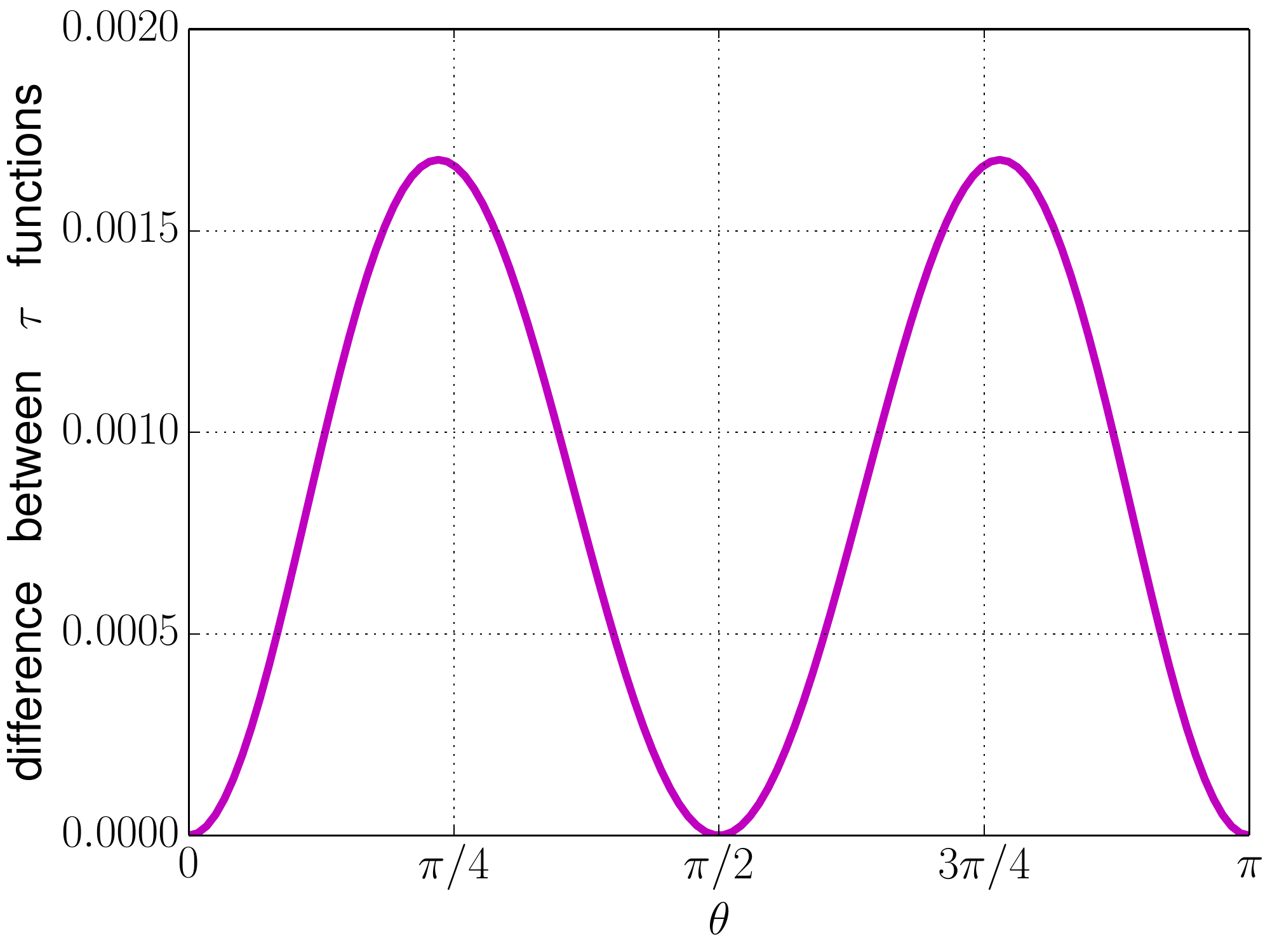}
    \caption{Difference of the proper times.}\label{f10N}
  \end{minipage}
\end{figure}

In order to obtain a more geometric and detailed comparison of
the two gauges, we consider the Eikonal equation following~\cite{frauendiener1998:_numer_hivp_ii}
\begin{equation}\label{eq:eiconal}
\nabla_a \tau \nabla^a \tau = -1.  
\end{equation}
Let $\tau$ be a smooth solution of the initial value problem of the
Eikonal equation with smooth initial data
$\tau_0:\Sigma_0\rightarrow\R$ prescribed freely on any smooth Cauchy
surface $\Sigma_0$ in any smooth globally hyperbolic spacetime. The
method of characteristics applied to this PDE allows to prove that
such a solution indeed always exists at least sufficiently close to
the initial hypersurface $\Sigma_0$.  For definiteness now we restrict
to the case of zero initial data $\tau_0=0$ for all of what
follows. Fix any point $p$ in the timelike future of $\Sigma_0$ in the
spacetime and consider any timelike geodesic through $p$ (with unit
tangent vector). Any such geodesic must intersect $\Sigma_0$ at some
point $x_0$ in the past of $p$. There is precisely one such timelike
geodesic through $p$ with unit tangent vector which intersects $\Sigma_0$ perpendicularly in $x_0$  and hence the point
$x_0$ is uniquely determined. 
The value
$\tau(p)$ of the solution $\tau$ of the Eikonal equation with zero
initial data then represents the proper time along this timelike
geodesic from $x_0$ to $p$.  The quantity $\tau$ is therefore a
meaningful geometric scalar quantity which can be used to compare our
numerical spacetimes, in particular when the same spacetime is
calculated in different coordinate gauges. We proceed
as follows. For initial data parameters  $R_0 = 2$, $c_1 = 1$ and $c_3 = 0.1$
  (see \Sectionref{sec:exactsol}):
\begin{enumerate}
\item We calculate the corresponding
  solution of Einstein's evolution equations in areal gauge
  (in the same way as in \Sectionref{sec:numareal}) and of
  the Eikonal equation \Eqref{eq:eiconal} (with zero initial data) up
  to $t=3$.  The value of the resulting $\tau$ function on the
  $t=3$-surface expressed with respect to spatial areal coordinates
  yields the dashed curve in \Figref{f9N}.
\item \Eqref{time_trans} implies that $t=3$ corresponds to
  $t_{(w)} \approx 4.217$. For the same initial data parameters as in
  the first step, we calculate the corresponding solution of
  Einstein's evolution equations in wave gauge numerically (using
  the gauge driver condition \Eqref{eq:gaugedriver} with $q=10$) and
  of the Eikonal equation \Eqref{eq:eiconal} (with zero initial data)
  up to $t_{(w)} \approx 4.217$.  The value of the resulting $\tau$
  function on the $t_{(w)} \approx 4.217$-surface expressed with
  respect to spatial wave map coordinates yields the continuous curve
  in \Figref{f9N}.
\end{enumerate}
Since the $t=3$-surface and the $t_{(w)} \approx 4.217$-surface
represent the same geometric surface in our spacetime and since $\tau$
is a geometric scalar quantity, the value of the solution of the
Eikonal equation on this surface should be the same function in both
steps above. However, since this function is expressed in terms of
different spatial coordinates, namely areal coordinates in the first
step and wave map coordinates in the second step, the two curves in \Figref{f9N} are slightly different. Hence
\Figref{f9N} can be understood as a representation of the difference
of these two sets of spatial coordinates. This difference is
emphasized in \Figref{f10N} where the two curves in
\Figref{f9N} are subtracted directly. Intuitively, these two sets of
spatial coordinates should agree at geometrically distinct points,
namely, at the poles and also at the equator as a consequence of a
reflection symmetry which is inherent to our particular class of exact
solutions. Indeed the difference curve in
\Figref{f10N} is zero at the poles $\theta=0,\pi$ and the equator
$\theta=\pi/2$.

\begin{figure}[t]
    \centering
    \includegraphics[width=0.49\linewidth]{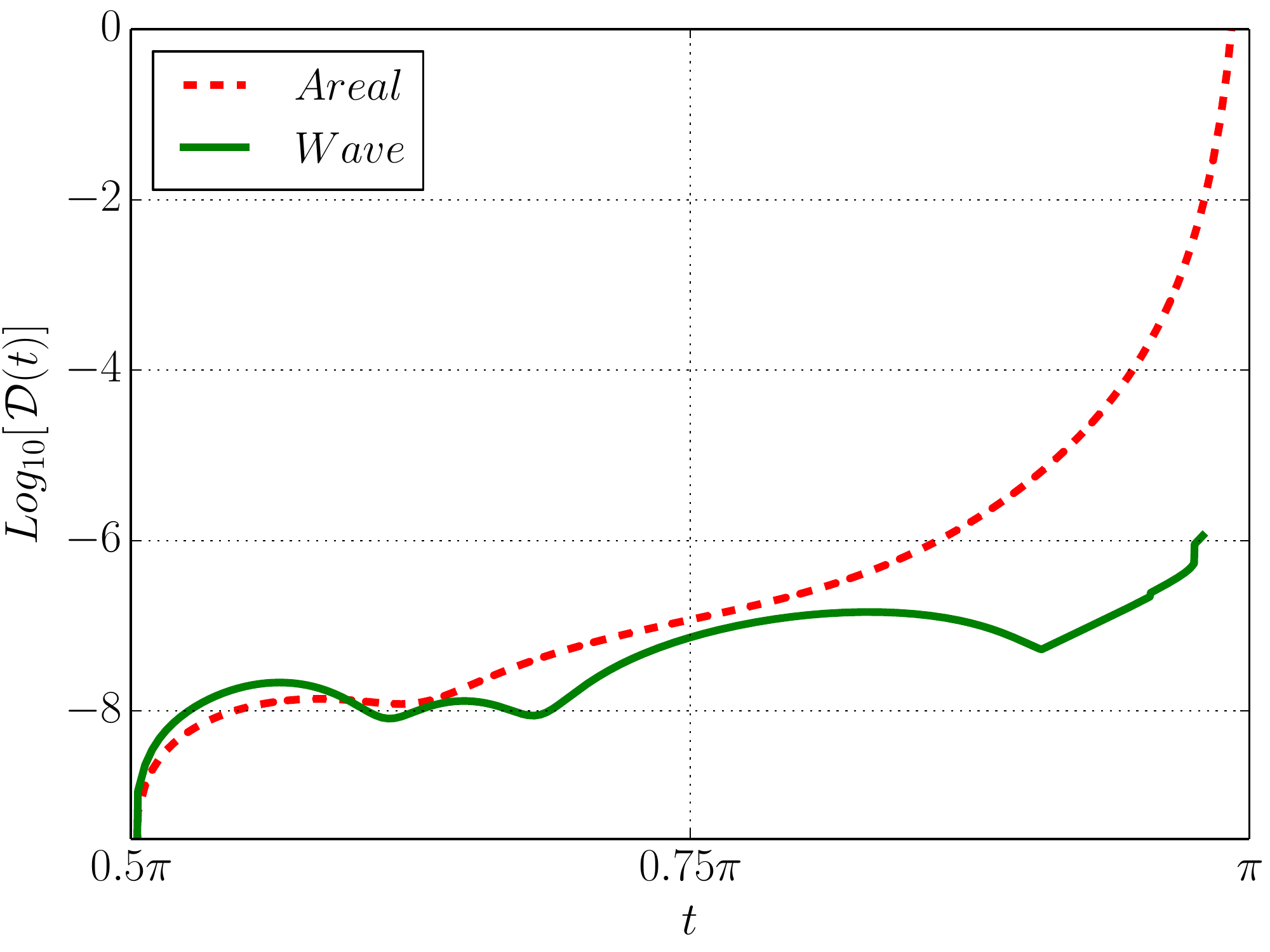} 
    \caption{Comparison of the constraint violations.}\label{f7N}   
\end{figure}
Next, we present plots of the constraint violations in both gauges; see
\Figref{f7N}. The dashed curve has been calculated in areal gauge (in the first
step above). The continuous curve has been calculated in wave map gauge (in the
second step above), but has then been expressed in terms of the areal time
function by means of \Eqref{time_trans}. It is interesting to notice that the
constraint violations are significantly smaller in wave map gauge than they are
in areal gauge towards the end of the numerical evolution.

Finally we comment on the fact that in wave map gauge the shift
quantity $\beta$ in \Eqref{3Dmetric_smoothframe} is a non-trivial
non-zero function in contrast to areal gauge; see
\Eqref{eq:beta}. When $\beta$ cannot be assumed to be zero
identically, the algebraic complexity of the evolution equations is
increased dramatically. It is surprising that irrespective of this it
appears that we get better numerical results in wave map than in
areal gauge.


\section{Discussion}
\label{sec:discussion}

The purpose of our work here was to introduce a numerical approach to
solve the Cauchy problem for spacetimes which involve the manifold
\St. We employ a fully regular representation of
the Einstein equations based on the spin-weighted spherical transform and the generalized wave map formalism. This allows us to account for all singular
terms explicitly which usually arise as a consequence of the coordinate
singularities of polar coordinates at the poles of the $2$-sphere.
Our numerical infrastructure is based on the spin-weight formalism and
corresponding transforms introduced in \cite{Beyer:2014bu,Beyer:2015bv}. We have
extended this infrastructure so that it now provides an efficient treatment of
axially symmetric functions on the $2$-sphere, reducing the complexity
$\mathcal{O}(L^3)$ of the full transform to the complexity
$\mathcal{O}(L^2)$. We therefore expect this method to be useful also for other
applications in future work.
We have also demonstrated the consistency and feasibility of our approach by means
of numerical studies of certain inhomogeneous cosmological solutions
of the Einstein's equations.

As another application of this method we are currently studying the critical
behavior of perturbations of the Nariai spacetime
\cite{Beyer:2014bu,Beyer:2015bv}.  In particular it is suggested that
larger amplitudes of the perturbations, which had not been studied
before, could lead to the formation of cosmological black holes. It
would be of great interest to explore the threshold solutions and the
expected cosmological black hole solutions as well as consequences for
the longstanding cosmic no-hair and cosmic censorship
conjectures. Other conceivable interesting applications where our numerical infrastructure can be applied directly are Robinson-Trautman solutions \cite{deOliveira:2011cm} and Ricci
flow \cite{Morgan:2007vs}.


\appendix

\section{The generalized wave map formalism}
\label{sec:generalized-wave-map}

Whether we want to solve the Cauchy problem for the $3+1$-Einstein vacuum equations $G_{ab}+\Lambda g_{ab}=0$ or for the $2+1$ equations \eqref{eq:geroch_einstein_equations}, the first task is always to extract hyperbolic evolution equations and constraint equations with well-understood propagation properties from the equation for the Ricci tensor of the unknown metric.
We shall now briefly discuss the ``generalized wave map formalism''. In most of the literature,
the related (but not covariant) generalized wave/harmonic formalism is used. While this is sufficient for
many applications, it is a drawback for us. In fact, for applications with
spatial $\St$-topologies covered by a single singular polar coordinate system,
it is far more convenient to work with actual covariant quantities (i.e., smooth
tensor fields). The reason is that frame components of smooth tensor fields on
$\St$ have well-defined spin-weights (despite of the fact that the frame itself is singular at the poles) so that they are expandable in
spin-weighted spherical harmonics, which are globally defined regular
`functions' on the $2$-sphere (even though their coordinate representation may be
singular). It turns out that expressing everything with respect to these bases, renders
the equations manifestly regular. 

To this end, we discuss the geometric formulation of the wave map gauge \cite{Friedrich:1991nn}. We consider a map $\Phi: M\to \bar M$ between two general smooth
4-dimensional manifolds $M$ and $\bar M$ (or open subsets thereof) equipped with
Lorentzian metrics\footnote{All of the following arguments also hold if $M$ and $\bar M$ are $n$-dimensional manifolds for some arbitrary positive integer $n$ and if $\bar h_{ab}$ is a general smooth pseudo-Riemannian (not necessary Lorentzian) metric.} $h_{ab}$ and $\bar h_{ab}$. The map $\Phi$ is called a wave map if it
extremizes the functional
\[
\mathcal{F}[\Phi] = \int_M \mathrm{tr}_h (\Phi^*\bar h)\,\mathrm{Vol}_h.
\]
In coordinate charts $(x^\mu)$ on $M$ and $(y^\alpha)$ on $\bar M$ we obtain the
Euler-Lagrange equations for the coordinate representation $y^\alpha =
y^\alpha(x^\mu)$ of $\Phi$
\begin{equation}
\Box_h y^\alpha + \bar\Gamma^\alpha{}_{\beta\gamma} h^{\mu\nu}\frac{\partial y^\beta}{\partial x^\mu}
\frac{\partial y^\gamma}{\partial x^\nu} = 0.\label{eq:wavemap}
\end{equation}
Here, $\bar\Gamma^\alpha{}_{\beta\gamma}$ are the Christoffel symbols for the
metric $\bar h_{ab}$ in the
coordinate basis on $\bar M$, and $\Box_h$ is the wave operator for scalar functions defined by $h_{ab}$. This equation is called the \emph{wave-map equation}. More details can be found in \cite{ChoquetBruhat:2008te}. If the
manifolds were Riemannian then the analogous equation would characterize a
harmonic map between $M$ and $\bar M$. Let us point out that the left hand side
of the equation defines a geometric object, namely a section in the pull-back
bundle $\Phi^*T\bar M$. This is not immediately obvious due to the appearance of
the Christoffel symbols in the second term. However, the tensorial character of
that term under change of coordinates in $\bar M$ is compensated for by the
first term which, by itself, is also non-tensorial under such coordinate
transformations.

The \emph{generalized wave-map} equation is the equation~\eqref{eq:wavemap} with
a non-vanishing, arbitrary right hand side, a section in $\Phi^*T \bar M$ with
coordinate representation $f^\alpha$
\begin{equation}
\Box_h y^\alpha + \bar\Gamma^\alpha{}_{\beta\gamma} h^{\mu\nu}\frac{\partial y^\beta}{\partial x^\mu}
\frac{\partial y^\gamma}{\partial x^\nu} = -f^\alpha.\label{eq:genwavemap}
\end{equation}
The minus sign on the right-hand side is a matter of conventions.
Suppose now that $\bar M = M$ and $\Phi = id_M$. Then $(x^\mu)$ and $(y^\alpha)$
are two coordinate charts for $M$ and~\eqref{eq:genwavemap} can be read as an
equation determining the coordinate system $(y^\alpha)$ for $M$ by imposing a geometrical
gauge condition. This equation is a semi-linear wave equation of $M$ which has solutions
near any Cauchy surface so that such a coordinate gauge always exists locally. 

Choosing the coordinates according to this gauge, i.e., putting $x^\mu = y^\mu$
and expressing the wave operator in these coordinates yields the equation
\[
(-\Gamma^\alpha{}_{\beta\gamma}  + \bar\Gamma^\alpha{}_{\beta\gamma})
h^{\beta\gamma} = -f^\alpha,
\]
where $\Gamma^\alpha{}_{\beta\gamma}$ are the Christoffel symbols of the metric
$h$ on $M$. In this equation the tensorial character becomes manifest since the
left hand side involves the difference of two connection coefficients so it
gives the components of a vector field in the coordinate basis of the
$(x^\mu)$. Therefore, this equation holds in any basis on $M$ as long as we
interpret the Christoffel symbols as the connection coefficients with
respect to the chosen basis. Note also that this implies that imposing the
equation~\eqref{eq:genwavemap} \emph{does not} constitute a condition on the
coordinate system $(x^\alpha)$ but a condition \emph{on the metric components} in their
dependence on the coordinates. We define the vector field $\mathcal{D}^a$ in
terms of its components
\begin{equation}
\label{eq:defDs}
\mathcal{D}^\alpha := 
(-\Gamma^\alpha{}_{\beta\gamma}  + \bar\Gamma^\alpha{}_{\beta\gamma})
h^{\beta\gamma} + f^\alpha.
\end{equation}
So, $\mathcal{D}^a = 0$ when \eqref{eq:genwavemap} is imposed. A metric $h_{ab}$
which is restricted by $\mathcal{D}^a = 0$ is said to be in wave map gauge (with
respect to $\bar{h}_{ab}$); in \Eqref{eq:background_metric} we fix a particular metric $\bar{h}_{ab}$. We point out that the wave map gauge reduces to the
widely used generalized wave/harmonic gauge characterized by $\Box x^\mu =- f^\mu$ on
space-times with topology $\mathbb{R}^4$ when the Minkowski metric in Cartesian
coordinates $x^\mu$ is used as a reference metric~$\bar{h}_{ab}$.


\begin{thebibliography}{10}

\bibitem{Ames:2012vz}
E.~Ames, F.~Beyer, J.~Isenberg, and P.~G. LeFloch.
\newblock {Quasilinear Hyperbolic Fuchsian Systems and AVTD Behavior in
  $T^2$-Symmetric Vacuum Spacetimes}.
\newblock {\em Ann. Henri Poincar{\'e}}, 14(6):1445--1523, 2013.

\bibitem{Beyer:2009vw}
F.~Beyer.
\newblock {A spectral solver for evolution problems with spatial $\mathbb
  S^3$-topology}.
\newblock {\em J. Comp. Phys.}, 228(17):6496--6513, 2009.

\bibitem{Beyer:2015bv}
F.~Beyer, B.~Daszuta, and J.~Frauendiener.
\newblock {A spectral method for half-integer spin fields based on
  spin-weighted spherical harmonics}.
\newblock {\em Class. Quantum Grav.}, 32(17):175013, 2015.

\bibitem{Beyer:2014bu}
F.~Beyer, B.~Daszuta, J.~Frauendiener, and B.~Whale.
\newblock {Numerical evolutions of fields on the 2-sphere using a spectral
  method based on spin-weighted spherical harmonics}.
\newblock {\em Class. Quantum Grav.}, 31(7):075019, 2014.

\bibitem{Beyer:2013uq}
F.~Beyer and L.~Escobar.
\newblock {Graceful exit from inflation for minimally coupled Bianchi A scalar
  field models}.
\newblock {\em Class. Quantum Grav.}, 30(19):195020, 2013.

\bibitem{Beyer:2011uz}
F.~Beyer and J.~Hennig.
\newblock {Smooth Gowdy-symmetric generalized Taub{\textendash}NUT solutions}.
\newblock {\em Class. Quantum Grav.}, 29(24):245017, 2012.

\bibitem{Beyer:2014vw}
F.~Beyer and J.~Hennig.
\newblock {An exact smooth Gowdy-symmetric generalized Taub{\textendash}NUT
  solution}.
\newblock {\em Class. Quantum Grav.}, 31(9):095010, 2014.

\bibitem{Boyle:2007kb}
M.~Boyle, L.~Lindblom, H.~P. Pfeiffer, M.~A. Scheel, and L.~E. Kidder.
\newblock {Testing the accuracy and stability of spectral methods in numerical
  relativity}.
\newblock {\em Phys. Rev. D}, 75(2):024006, 2007.

\bibitem{Brodbeck:1999en}
O.~Brodbeck, S.~Frittelli, P.~H{\"u}bner, and O.~A. Reula.
\newblock {Einstein{\textquoteright}s equations with asymptotically stable
  constraint propagation}.
\newblock {\em J. Math. Phys.}, 40(2):909, 1999.

\bibitem{Brugmann:2013kt}
B.~Br{\"u}gmann.
\newblock {A pseudospectral matrix method for time-dependent tensor fields on a
  spherical shell}.
\newblock {\em J. Comp. Phys.}, 235:216--240, 2013.

\bibitem{choptuik2003axisymmetric}
M.~W. Choptuik, E.~W. Hirschmann, S.~L. Liebling, and F.~Pretorius.
\newblock {An axisymmetric gravitational collapse code}.
\newblock {\em Class. Quantum Grav.}, 20(9):1857, 2003.

\bibitem{ChoquetBruhat:2008te}
Y.~Choquet-Bruhat.
\newblock {\em {General Relativity and the Einstein Equations}}.
\newblock Oxford mathematical monographs. Oxford University Press, Oxford, New
  York, 2008.

\bibitem{ChoquetBruhat:1969cl}
Y.~Choquet-Bruhat and R.~P. Geroch.
\newblock {Global aspects of the Cauchy problem in general relativity}.
\newblock {\em Commun. Math. Phys.}, 14(4):329--335, 1969.

\bibitem{Chrusciel:1990ti}
P.~T. Chru{\'{s}}ciel.
\newblock {On space-times with $U(1)\times U(1)$ symmetric compact Cauchy
  surfaces}.
\newblock {\em Ann. Phys.}, 202(1):100--150, 1990.

\bibitem{deOliveira:2011cm}
H.~P. de~Oliveira, E.~L. Rodrigues, and J.~E.~F. Skea.
\newblock {Numerical evolution of general Robinson-Trautman spacetimes: Code
  tests, wave forms, and the efficiency of the gravitational wave extraction}.
\newblock {\em Phys. Rev. D}, 84(4):044007, 2011.

\bibitem{Durran:2010}
D.~R. Durran.
\newblock {\em {Numerical Methods for Fluid Dynamics}}, volume~32 of {\em Texts
  in Applied Mathematics}.
\newblock Springer, New York, NY, 2010.

\bibitem{Fornberg:1998gv}
B.~Fornberg.
\newblock {\em {A Practical Guide to Pseudospectral Methods}}.
\newblock Cambridge University Press, 1998.

\bibitem{FouresBruhat:1952ji}
Y.~Four{\`e}s-Bruhat.
\newblock {Th{\'e}or{\`e}me d'existence pour certains syst{\`e}mes
  d'{\'e}quations aux d{\'e}riv{\'e}es partielles non lin{\'e}aires}.
\newblock {\em Acta Mathematica}, 88(1):141--225, 1952.

\bibitem{frauendiener1998:_numer_hivp_ii}
J.~Frauendiener.
\newblock {Numerical treatment of the hyperboloidal initial value problem for
  the vacuum Einstein equations. II. The evolution equations}.
\newblock {\em Phys. Rev. D}, 58(6):064003, 1998.

\bibitem{Frauendiener04}
J.~Frauendiener and T.~Vogel.
\newblock {Algebraic stability analysis of constraint propagation}.
\newblock {\em Class. Quantum Grav.}, 22(9):1769--1793, 2005.

\bibitem{Friedrich:1991nn}
H.~Friedrich.
\newblock {On the global existence and the asymptotic behavior of solutions to
  the Einstein-Maxwell-Yang-Mills equations}.
\newblock {\em J. Diff. Geom.}, 34:275--345, 1991.

\bibitem{Garfinkle:1999ix}
D.~Garfinkle.
\newblock {Numerical simulations of Gowdy spacetimes on $S^2\times S^1\times
  R$}.
\newblock {\em Phys. Rev. D}, 60(10):104010, 1999.

\bibitem{Garfinkle:2002ft}
D.~Garfinkle.
\newblock {Harmonic coordinate method for simulating generic singularities}.
\newblock {\em Phys. Rev. D}, 65(4):044029, 2002.

\bibitem{Geroch:1971ix}
R.~P. Geroch.
\newblock {A Method for Generating Solutions of Einstein's Equations}.
\newblock {\em J. Math. Phys.}, 12(6):918, 1971.

\bibitem{gomez1997eth}
R.~G{\'o}mez, L.~Lehner, P.~Papadopoulos, and J.~Winicour.
\newblock {The eth formalism in numerical relativity}.
\newblock {\em Class. Quantum Grav.}, 14(4):977--990, 1997.

\bibitem{Gundlach05}
C.~Gundlach, J.~Mart{\'\i}n-Garc{\'\i}a, G.~Calabrese, and I.~Hinder.
\newblock {Constraint damping in the Z4 formulation and harmonic gauge}.
\newblock {\em Class. Quantum Grav.}, 22(17):3767--3774, 2005.

\bibitem{Hennig:2013eu}
J.~Hennig.
\newblock {Fully pseudospectral time evolution and its application to $1+1$
  dimensional physical problems}.
\newblock {\em J. Comp. Phys.}, 235:322--333, 2013.

\bibitem{Holst:2004ep}
M.~Holst, L.~Lindblom, R.~Owen, H.~P. Pfeiffer, M.~A. Scheel, and L.~E. Kidder.
\newblock {Optimal constraint projection for hyperbolic evolution systems}.
\newblock {\em Phys. Rev. D}, 70(8):084017, 2004.

\bibitem{Huffenberger:2010hh}
K.~M. Huffenberger and B.~D. Wandelt.
\newblock {Fast and Exact Spin-s Spherical Harmonic Transforms}.
\newblock {\em Astro. J. Suppl. Series}, 189(2):255--260, 2010.

\bibitem{Lehner:2005hc}
L.~Lehner, O.~A. Reula, and M.~Tiglio.
\newblock {Multi-block simulations in general relativity: high-order
  discretizations, numerical stability and applications}.
\newblock {\em Class. Quantum Grav.}, 22(24):5283--5321, 2005.

\bibitem{Lindblom:2004ci}
L.~Lindblom, M.~A. Scheel, L.~E. Kidder, H.~P. Pfeiffer, D.~Shoemaker, and
  S.~A. Teukolsky.
\newblock {Controlling the growth of constraints in hyperbolic evolution
  systems}.
\newblock {\em Phys. Rev. D}, 69(12):124025, 2004.

\bibitem{Lindblom:2009in}
L.~Lindblom and B.~Szilagyi.
\newblock {Improved gauge driver for the generalized harmonic Einstein system}.
\newblock {\em Phys. Rev. D}, 80(8):084019, 2009.

\bibitem{Moncrief:1984js}
V.~Moncrief.
\newblock {The space of (generalized) Taub-Nut spacetimes}.
\newblock {\em J. Geom. Phys.}, 1(1):107--130, 1984.

\bibitem{Moncrief:1986th}
V.~Moncrief.
\newblock {Reduction of Einstein's equations for vacuum space-times with
  spacelike $U(1)$ isometry groups}.
\newblock {\em Ann. Phys.}, 167(1):118--142, 1986.

\bibitem{Morgan:2007vs}
J.~Morgan and T.~Gang.
\newblock {\em {Ricci Flow and the Poincar{\'e} Conjecture}}.
\newblock Clay Mathematics Monographs, 2007.

\bibitem{newman1966note}
E.~T. Newman and R.~Penrose.
\newblock {Note on the Bondi-Metzner-Sachs Group}.
\newblock {\em J. Math. Phys.}, 7(5):863, 1966.

\bibitem{Newman:1963up}
E.~T. Newman, L.~Tamburino, and T.~Unti.
\newblock {Empty-Space Generalization of the Schwarzschild Metric}.
\newblock {\em J. Math. Phys.}, 4(7):915, 1963.

\bibitem{Preto2005:Evolution-of-Binary-Black-Hole}
F.~Pretorius.
\newblock {Evolution of Binary Black-Hole Spacetimes}.
\newblock {\em Phys. Rev. Lett.}, 95:121101, 2005.

\bibitem{Ringstrom:2009cj}
H.~Ringstr{\"o}m.
\newblock {\em {The Cauchy Problem in General Relativity}}.
\newblock ESI Lectures in Mathematics and Physics. European Mathematical
  Society, Z{\"u}rich, Switzerland, 2009.

\bibitem{Rinne:2010ew}
O.~Rinne.
\newblock {An axisymmetric evolution code for the Einstein equations on
  hyperboloidal slices}.
\newblock {\em Class. Quantum Grav.}, 27(3):035014, 2010.

\bibitem{Risbo:1996iy}
T.~Risbo.
\newblock {Fourier transform summation of Legendre series and D-functions}.
\newblock {\em J. Geodesy}, 70(7):383--396, 1996.

\bibitem{Stephani:2003uy}
H.~Stephani, D.~Kramer, M.~A.~H. MacCallum, C.~Hoenselaers, and E.~Herlt.
\newblock {\em {Exact solutions of Einstein's field equations}}.
\newblock Cambridge University Press, second edition, 2003.

\bibitem{Sugiura:1990vj}
M.~Sugiura.
\newblock {\em {Unitary Representations and Harmonic Analysis}}.
\newblock An Introduction. North Holland, 1990.

\bibitem{Taub:1951vk}
A.~H. Taub.
\newblock {Empty Space-Times Admitting a Three Parameter Group of Motions}.
\newblock {\em Ann. Math.}, 53(3):472--490, 1951.

\bibitem{Trapani:2006he}
S.~Trapani and J.~Navaza.
\newblock {Calculation of spherical harmonics and Wigner d-functions by FFT.
  Applications to fast rotational matching in molecular replacement and
  implementation into AMoRe}.
\newblock {\em Acta Cryst. A}, 62(4):262--269, 2006.

\bibitem{Vretblad:2003}
A.~Vretblad.
\newblock {\em {Fourier Analysis and Its Applications}}.
\newblock Springer New York, 2003.

\end{thebibliography}

\end{document}